\documentclass[a4paper,fleqn,usenatbib]{mnras}

\usepackage{mathptmx}
\usepackage{txfonts}

\usepackage[T1]{fontenc}
\usepackage{ae,aecompl}

\usepackage{graphicx}	
\usepackage{amssymb}	

\usepackage{amsfonts}
\usepackage{calrsfs}
\usepackage{txfonts}
\usepackage{scrextend}
\usepackage[hyphenbreaks]{breakurl}

\usepackage{caption}
\usepackage{subcaption}
\usepackage{threeparttable}

\usepackage{longtable}

\title[Follow-up of TRAPPIST-1 with \textit{Spitzer}]{Early 2017 observations of TRAPPIST-1 with \textit{Spitzer}}

\author[Delrez et al.]{L.~Delrez,$^{1}$\thanks{E-mail: \url{lcd44@cam.ac.uk}}
M.~Gillon,$^{2}$
A. H. M. J.~Triaud,$^{3}$
B.-O.~Demory,$^{4}$
J.~de Wit,$^{5}$
\newauthor
J. G.~Ingalls,$^{6}$
E. Agol,$^{7,8}$
E. Bolmont,$^{9,10}$ 
A.~Burdanov,$^{2}$
A.~J.~Burgasser,$^{11}$
\newauthor 
S. J.~Carey,$^{6}$
E.~Jehin,$^{2}$
J. Leconte,$^{12}$
S. Lederer,$^{13}$  
D.~Queloz,$^{1}$
F. Selsis,$^{12}$ 
\newauthor 
and V.~Van Grootel$^{2}$  
\vspace{0.1cm}
\\
$^1$ Cavendish Laboratory, JJ Thomson Avenue, Cambridge, CB3 0HE, UK\\
$^2$ Space Sciences, Technologies and Astrophysics Research (STAR) Institute, Universit\'e de Li\`ege, All\'ee du 6 Ao\^ut 19C,\\ B-4000 Li\`ege, Belgium\\
$^3$ School of Physics \& Astronomy, University of Birmingham, Edgbaston, Birmingham B15 2TT, United Kingdom\\
$^4$ University of Bern, Center for Space and Habitability, Gesellschaftsstrasse 6, CH-3012, Bern, Switzerland\\
$^5$ Department of Earth, Atmospheric and Planetary Sciences, Massachusetts Institute of Technology, 77 Massachusetts Avenue,\\ Cambridge, MA 02139, USA\\
$^6$ IPAC, California Institute of Technology, 1200 E California Boulevard, Mail Code 314-6, Pasadena, California 91125, USA\\
$^7$ Astronomy Department, University of Washington, Seattle, WA, 98195, USA; Guggenheim Fellow\\   
$^8$ NASA Astrobiology Institute's Virtual Planetary Laboratory, Seattle, WA, 98195, USA\\
$^{9}$ IRFU, CEA, Universit\'e Paris-Saclay, F-91191 Gif-sur-Yvette, France\\ 
$^{10}$ Universit\'e Paris Diderot, AIM, Sorbonne Paris Cit\'e, CEA, CNRS, F-91191 Gif-sur-Yvette, France\\
$^{11}$ Center for Astrophysics and Space Science, University of California San Diego, La Jolla, CA, 92093, USA\\
$^{12}$ Laboratoire d'astrophysique de Bordeaux, Univ. Bordeaux, CNRS, B18N, All\'ee Geoffroy Saint-Hilaire, F-33615 Pessac, France\\
$^{13}$ NASA Johnson Space Center, 2101 NASA Parkway, Houston, Texas, 77058, USA\\
}

\date{Accepted 2017 December 31. Received 2017 December 21; in original form 2017 October 12}

\pubyear{2017}

\begin{document}
\label{firstpage}
\pagerange{\pageref{firstpage}--\pageref{lastpage}}
\maketitle

\begin{abstract}
The recently detected TRAPPIST-1 planetary system, with its seven planets transiting a nearby ultracool dwarf star, offers the first opportunity to perform comparative exoplanetology of temperate Earth-sized worlds. To further advance our understanding of these planets' compositions, energy budgets, and dynamics, we are carrying out an intensive photometric monitoring campaign of their transits with the \textit{Spitzer Space Telescope}. In this context, we present 60 new transits of the TRAPPIST-1 planets observed with \textit{Spitzer}/IRAC in February and March 2017. We combine these observations with previously published \textit{Spitzer} transit photometry and perform a global analysis of the resulting extensive dataset. This analysis refines the transit parameters and provides revised values for the planets' physical parameters, notably their radii, using updated properties for the star. As part of our study, we also measure precise transit timings that will be used in a companion paper to refine the planets' masses and compositions using the transit timing variations method. TRAPPIST-1 shows a very low level of low-frequency variability in the IRAC 4.5-$\mu$m band, with a photometric RMS of only 0.11\% at a 123-s cadence. We do not detect any evidence of a (quasi-)periodic signal related to stellar rotation. We also analyze the transit light curves individually, to search for possible variations in the transit parameters of each planet due to stellar variability, and find that the \textit{Spitzer} transits of the planets are mostly immune to the effects of stellar variations. These results are encouraging for forthcoming transmission spectroscopy observations of the TRAPPIST-1 planets with the \textit{James Webb Space Telescope}.
\vspace{0.3cm}
\end{abstract}

\begin{keywords}
planetary systems -- stars: individual: TRAPPIST-1 -- techniques: photometric
\vspace{1cm}
\end{keywords}


\defcitealias{Gillon2016}{G16}
\defcitealias{Gillon2017}{G17}


\section{Introduction}

Small stars are beneficial for the discovery and study of exoplanets by transit methods \citep[e.g.][]{Nutzman2008,He2017}. For a given planet's size and irradiation, they offer deeper planetary transits, and shorter orbital periods. The seven Earth-sized worlds orbiting the nearby ultracool dwarf star TRAPPIST-1 \citep[][hereafter \citetalias{Gillon2016} and \citetalias{Gillon2017} respectively]{Gillon2016,Gillon2017} have become prime targets for the study of small planets beyond the Solar system, including the study of their atmospheres, owing to their transiting configuration combined with the infrared brightness ($K$=10.3) and Jupiter-like size \hbox{($\sim$0.12 $R_{\odot}$)} of their host star \citep{de-Wit2016,Barstow2016,Morley2017}.\\ 
\indent
The TRAPPIST-1 planets have further importance. There are approximately three times as many M-dwarfs as FGK-dwarfs in the Milky Way \citep{Kroupa2001,Chabrier2003,Henry2006}, and small planets appear to surround M-dwarfs three to five times more frequently than Sun-like stars \citep[e.g.][]{Bonfils2013,Dressing2013,Dressing2015}. If this trend continues to the bottom of the main-sequence \citep[see][]{He2017}, the TRAPPIST-1 planets could well represent the most common Earth-sized planets in our Galaxy, which in itself would warrant special attention. \hbox{TRAPPIST-1} also presents an interesting numerical and dynamical challenge; for example, assessing its stability on Gyr timescales for orbital periods that have day to week timescales \citep{Tamayo2016,Tamayo2017}.\\ 
\indent
A comparative study of the TRAPPIST-1 planets is aided by the fact that they all transit the same star. Because the knowledge of most planetary parameters (e.g., mass and radius) is dependent on knowing these parameters for their parent stars, it is often difficult to make accurate comparisons across systems. Although the parameters of the TRAPPIST-1 planets remain dependent on the parameters of their host star, we can nevertheless compare the planets to one another. Furthermore, the planets have emerged from the same nebular environment, have experienced a similar history in terms of irradiation \citep[notably in the XUV range;][]{Wheatley2017,Bourrier2017,OMalley-James2017}, and similar volatile delivery and atmospheric sculpting via cometary impacts (Kral et al. submitted). Thus, any differences among the planets must be the result of distinctions in their development. One example would be the possible detection of $O_2$, which on Earth has biological origins but on other worlds can be produced abiologically through the photodissociation of water vapour and escape of hydrogen \citep[e.g.][]{Wordsworth2014, Bolmont2017}. The presence of significant $O_2$ on only one of the seven planets would indicate a process particular to that planet, such as microbial respiration, with potentially far-reaching implications in humanity's search for life beyond the Earth.\\
\indent
To improve the characterization of the planets in the TRAPPIST-1 system, and prepare for exploration of their atmospheric properties with the upcoming {\it James Webb Space Telescope} (JWST) mission \citep{Barstow2016,Morley2017}, we are conducting an intensive, high-precision, space-based photometric monitoring campaign of the system with {\it Spitzer} (Exploration Program ID 13067). The main goals of this program are to improve the planets' transit parameters - notably to refine the determination of their radii - and to use the measured transit timing variations (TTVs) to constrain their masses and orbits \citep[][]{Agol:2005qy, Holman:2005fk}. Our {\it Spitzer} program also aims to study the infrared variability of the host star and its possible impact on the future JWST observations, and to obtain first constraints on the presence of atmospheres around the planets by comparing their transit depths measured in the infrared by {\it Spitzer} to the ones measured at shorter wavelengths by other facilities \citep[e.g.][]{de-Wit2016}.\\
\indent 
In this paper, we present observations gathered during the Feb-Mar 2017 window of visibility of the star by {\it Spitzer}. These new data more than double the number of transit events observed on TRAPPIST-1 with {\it Spitzer}. Section \ref{data_reduc} describes the data and data reduction. In Section \ref{analyses}, we combine these observations with previous {\it Spitzer} transit photometry of TRAPPIST-1 presented in \citetalias{Gillon2017}, and perform a global analysis that enables us to significantly improve the planets' transit parameters. We also use updated physical parameters for the star \citep[][]{VanGrootel2017}, to revise the planets' physical parameters, notably their radii. In addition, we assess the low-frequency infrared variability of the star and its impact on our measured quantities. As part of our analysis, we also extract precise transit timings. Our TTV analysis of the current timing dataset, including these new {\it Spitzer} timings, the resulting updated planetary masses, and our interpretations on the planets' composition and formation, are presented in a separate companion paper \citep{Grimm2017}. In Section \ref{discussion}, we discuss our results, examining variability in the transit parameters over the breadth of the dataset, and setting limits on wavelength-dependent transit depths for TRAPPIST-1b as a probe of its atmospheric properties. We summarize our results in \hbox{Section \ref{conclusion}}.


\section{Observations and data reduction}  
\label{data_reduc}

As part of our Warm \textit{Spitzer} Exploration Science program (ID 13067), \textit{Spitzer} monitored 9, 16, 9, 6, 4, 3, and 1 new transit(s) of TRAPPIST-1b, -1c, -1d, -1e, -1f, -1g, and \hbox{-1h}, respectively, in the 4.5-$\mu$m channel of its Infrared Array Camera (IRAC, \citealt{Fazio2004}). Twelve additional transits of \hbox{TRAPPIST-1b} were also observed with IRAC in the \hbox{3.6-$\mu$m} channel. All these observations were performed between \hbox{18 Feb} and 27 Mar 2017. The corresponding data can be accessed using the Spitzer Heritage Archive\footnote{\url{http://sha.ipac.caltech.edu}} (SHA). The observations were obtained in subarray mode (\hbox{32 $\times$ 32} pixels windowing of the detector) with an exposure time of \hbox{1.92 s}. They were made without dithering (continuous staring) and in the pointing calibration and reference sensor (PCRS) peak-up mode \citep{Ingalls2012}, which maximizes the accuracy in the position of the target on the detector so as to minimize the so-called ``pixel phase effect'' of IRAC indium antinomide arrays (e.g. \citealt{Knutson2008}). All the data were calibrated with the \textit{Spitzer} pipeline S19.2.0, and delivered as cubes of 64 subarray images of 32 $\times$ 32 pixels (pixel scale = 1.2 arcsec). \\
\indent
Our photometric extraction was identical to that described in \citetalias{Gillon2017}. After converting fluxes from MJy/sr to photon counts, we used the \textsc{iraf/daophot}\footnote{IRAF is distributed by the National Optical Astronomy Observatory, which is operated by the Association of Universities for Research in Astronomy, Inc., under cooperative agreement with the National Science Foundation.} software \citep{Stetson1987} to measure aperture photometry for the star within a circular aperture of 2.3 pixels. The apertures for each image were centered on the stellar point-spread function (PSF) by fitting to a 2D-Gaussian profile, which also yielded measurements of the PSF width along the $x$ and $y$ image coordinates. Images with discrepant measurements for the PSF center, background flux, or source flux were discarded as described in \cite{Gillon2014}. We then combined the measurements per cube of 64 images. The photometric errors were taken as the errors on the average flux measurements for each cube. 
\\
\indent
We complemented the resulting set of light curves with the \textit{Spitzer} transit photometry previously published in \citetalias{Gillon2017}, consisting of 16, 11, 5, 2, 3, 2, and 1 transit(s) of TRAPPIST-1b, -1c, -1d, -1e, -1f, -1g, and -1h, respectively, all observed with IRAC at 4.5 $\mu$m. We refer the reader to \citetalias{Gillon2017} and references therein for more details about these data.\\
\indent
Our extensive \textit{Spitzer} dataset thus includes a total of 37, 27, 14, 8, 7, 5, and 2 transits of planets b, c, d, e, f, g, and h, respectively. A few of these light curves showed flare signatures for which we discarded the corresponding measurements.


\section{Data analysis} 
\label{analyses}

Our data analysis was divided into three steps. We first performed individual analyses of the transit light curves (Section \ref{indiv_analyses}) to select the optimal photometric model for each light curve, measure the transit timings, and assess the variability of the transit parameters for each planet due to stellar variability. We then performed a global analysis of the whole dataset (Section \ref{global_analysis}) with the aim of improving our knowledge of the system parameters. Finally, we used our extensive \textit{Spitzer} dataset to assess the low-frequency infrared variability of the star in the IRAC 4.5-$\mu$m channel (Section \ref{overall_variability}).\\
\indent
Our individual and global data analyses were all carried out using the most recent version of the adaptive Markov Chain Monte-Carlo (MCMC) code presented in \citeauthor{Gillon2012} (\citeyear{Gillon2012}, see also \citealt{Gillon2014}). We refer the reader to these papers and references therein for a detailed description of the MCMC algorithm and only describe below the aspects that are specific to the analyses presented here. 

\subsection{Individual analyses}\label{sec:indi}  
\label{indiv_analyses}

We first converted each UT time of mid-exposure to the BJD$_{\mathrm{TDB}}$ time-scale, as described by \cite{Eastman2010}. We then performed an individual model selection for each light curve, using the transit model of \cite{Mandel2002} multiplied by a photometric baseline model, different for each light curve, aiming to represent the other astrophysical and instrumental effects at the source of photometric variations. A quadratic limb-darkening law was assumed for the star. For each light curve, we explored a large range of baseline models and selected the one that minimizes the Bayesian Information Criterion (BIC, \citealt{Schwarz1978}). This led us to first include a linear or quadratic function of the $x$- and $y$-positions of the stellar PSF center (as measured in the images by fitting a two-dimensional Gaussian profile) in the baseline model of every light curve to account for the pixel phase effect (e.g. \citealt{Knutson2008}). For some light curves, the modeling of this effect was improved by complementing the $x$- and $y$-polynomial with a numerical position model computed with the bi-linearly-interpolated sub-pixel sensitivity (BLISS) mapping method \citep{Stevenson2012}. The sampling of the position space (number of grid points) was selected so that at least five measurements were located in each sub-pixel box. We refer the reader to \cite{Gillon2014} for more details about the implementation of this approach in the MCMC code. We found that including a linear or quadratic function of the measured FWHM of the stellar PSF in the $x$- and/or $y$-directions resulted in a significant decrease in the BIC for most light curves (similar to \citealt{Lanotte2014} and \citealt{Demory2016}). A polynomial of the logarithm of time was also required for some light curves to represent a sharp decrease of the detector response at the beginning of the observations (the so-called ``ramp effect'', \citealt{Knutson2008}). Finally, a polynomial of time was also included in the baseline model for a fraction of the light curves to represent low-frequency signals likely related to the rotational variability of the star (see \citetalias{Gillon2016}, \citealt{Luger2017}, and Section \ref{overall_variability}).\\
\indent
For each individual analysis, the jump parameters of the MCMC, i.e. the parameters randomly perturbed at each step of the Markov chains, were as follows:
\begin{itemize}
    \item The stellar mass $M_{\star}$, radius $R_{\star}$, effective temperature $T_{\rm{eff}}$, and metallicity [Fe/H]. We assumed the normal distributions $\mathcal{N}(0.0802,0.0073^{2})$ $M_{\odot}$, $\mathcal{N}(0.117,0.0036^{2})$ $R_{\odot}$, $\mathcal{N}(2559,50^{2})$ K, and $\mathcal{N}(0.04,0.08^{2})$ dex as respective prior probability distribution functions (PDFs) for these four parameters based on the values given in \citetalias{Gillon2017}.
    \item For each transit (some light curves cover several transits), the transit depth \hbox{d$F$ = $(R_{\mathrm{p}}/R_{\star})^{2}$} where $R_{\mathrm{p}}$ is the radius of the transiting planet, the transit impact parameter \hbox{$b$ = $a$ $\mathrm{cos}$ $i/R_{\star}$} where $a$ is the orbital semi-major axis and $i$ is the orbital inclination, the transit width $W$ (defined as the duration from first to last contact), and the time of mid-transit $T_{0}$.
    \item The linear combinations of the quadratic limb-darkening coefficients ($u_{1}$, $u_{2}$) in the considered bandpass, $c_{1} = 2 \times u_{1} + u_{2}$ and $c_{2} = u_{1} - 2 \times u_{2}$. For each bandpass, values and errors for $u_{1}$ and $u_{2}$ were interpolated for TRAPPIST-1 from the tables of \cite{Claret2011} and the corresponding normal distributions were used as prior PDFs.
\end{itemize}
\noindent
 For these individual analyses, we kept the orbital period(s) of the relevant planet(s) fixed to the value(s) reported in \citetalias{Gillon2017} (for TRAPPIST-1b, c, d, e, f, g) and \citeauthor{Luger2017} (\citeyear{Luger2017}, for TRAPPIST-1h). As in \citetalias{Gillon2017}, we assumed circular orbits for all the planets (eccentricity $e=0$).\\
\begin{figure*}
\center
\begin{subfigure}[b]{0.49\textwidth}
	\includegraphics[width= \textwidth]{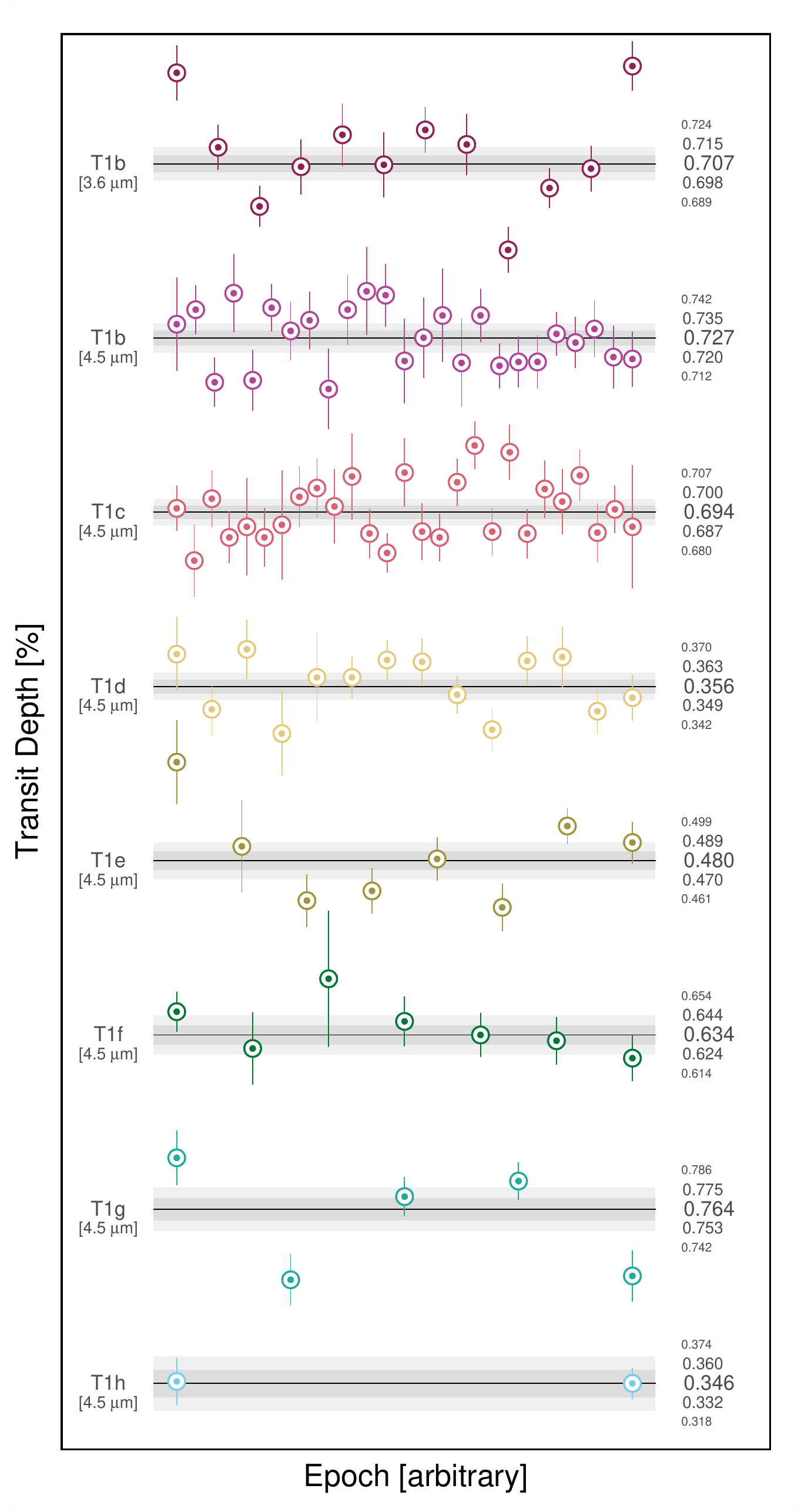}
\end{subfigure}
\begin{subfigure}[b]{0.49\textwidth}
	\includegraphics[width= \textwidth]{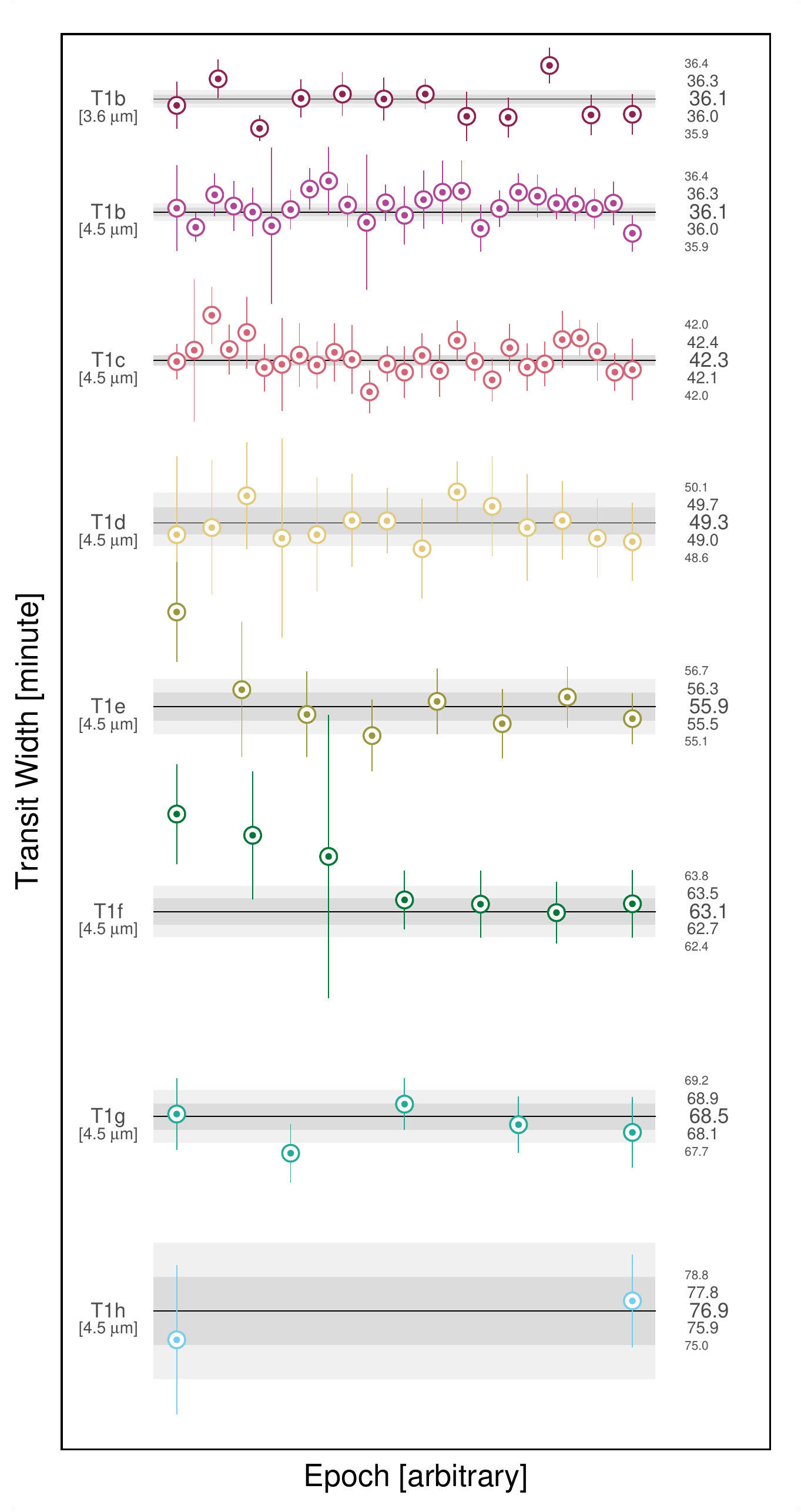}
\end{subfigure}
\caption{\textit{Left:} Individual transit depth measurements for each of the events captured with {\it Spitzer}. The horizontal black line shows the median of the global MCMC posterior PDF (with its 1, and 2$\sigma$ confidence, in shades of grey); the numeral values are also provided. Events are ranked in order of capture, left to right (but not linearly in time).
\textit{Right:} Similarly, but showing the duration of transit.
}\label{fig:singles}  
\vspace{0.3cm}
\end{figure*}  
\indent
For each light curve, a preliminary MCMC analysis composed of one chain of \hbox{10 000} steps was first performed to estimate the correction factors $CF$ to be applied to the photometric error bars, to account for both the over- or under-estimation of the white noise of each measurement and the presence of correlated (red) noise in the data (see \citealt{Gillon2012} and Appendix~\ref{plotrms} for details). Then, a longer MCMC analysis was performed, composed of two chains of \hbox{100 000} steps, whose convergence was checked using the statistical test of \cite{Gelman1992}.\\
\indent
Table \ref{table_indiv} presents for each planet the transit timings, depths, and durations deduced from the individual analyses of its transit light curves. For each planet, we performed a linear regression of the measured {\it Spitzer} transit timings as a function of their epochs to derive an updated mean transit ephemeris (given in Table \ref{table_glob}). We show individual depths and durations in Fig.~\ref{fig:singles} and see that in general, they are compatible to one another, epoch after epoch, following close to a normal distribution. Our individual uncertainties on the duration appear all slightly over-estimated when we compare them to the mean of individual measurements. All planets have reduced chi-squared $\chi^2_r <1$ except for TRAPPIST-1b at \hbox{3.6 $\mu$m}, which has $\chi^2_r = 1.1$. The situation is different for the transit depths. TRAPPIST-1b (4.5 $\mu$m), -1c, -1d, -1f, and -1h have $\chi^2_r$ compatible with normal distribution, whereas \hbox{TRAPPIST-1b} (3.6 $\mu$m) has $\chi^2_r = 4.4$, TRAPPIST-1e has $\chi^2_r = 2.4$, and TRAPPIST-1g has $\chi^2_r = 4.3$. We discuss these dispersions later in the text (Section \ref{sec:variability}).

\subsection{Global analysis}   
\label{global_analysis}

\begin{figure}
\centering
\includegraphics[width=0.49\textwidth]{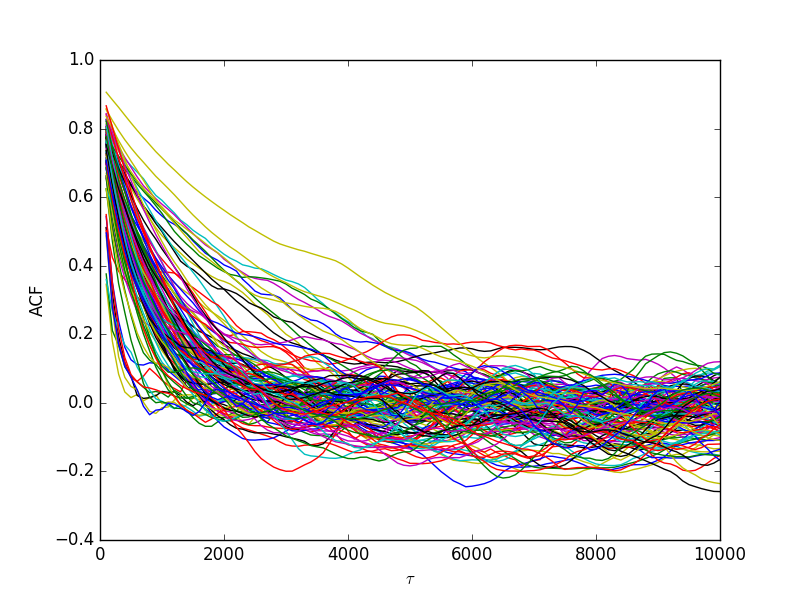}
\caption{Autocorrelation function for all 122 jump parameters versus chain lag, $\tau$.
}\label{fig:acf} 
\end{figure}

\begin{figure*}
\center
\begin{subfigure}[b]{0.505\textwidth}
	\includegraphics[width= \textwidth]{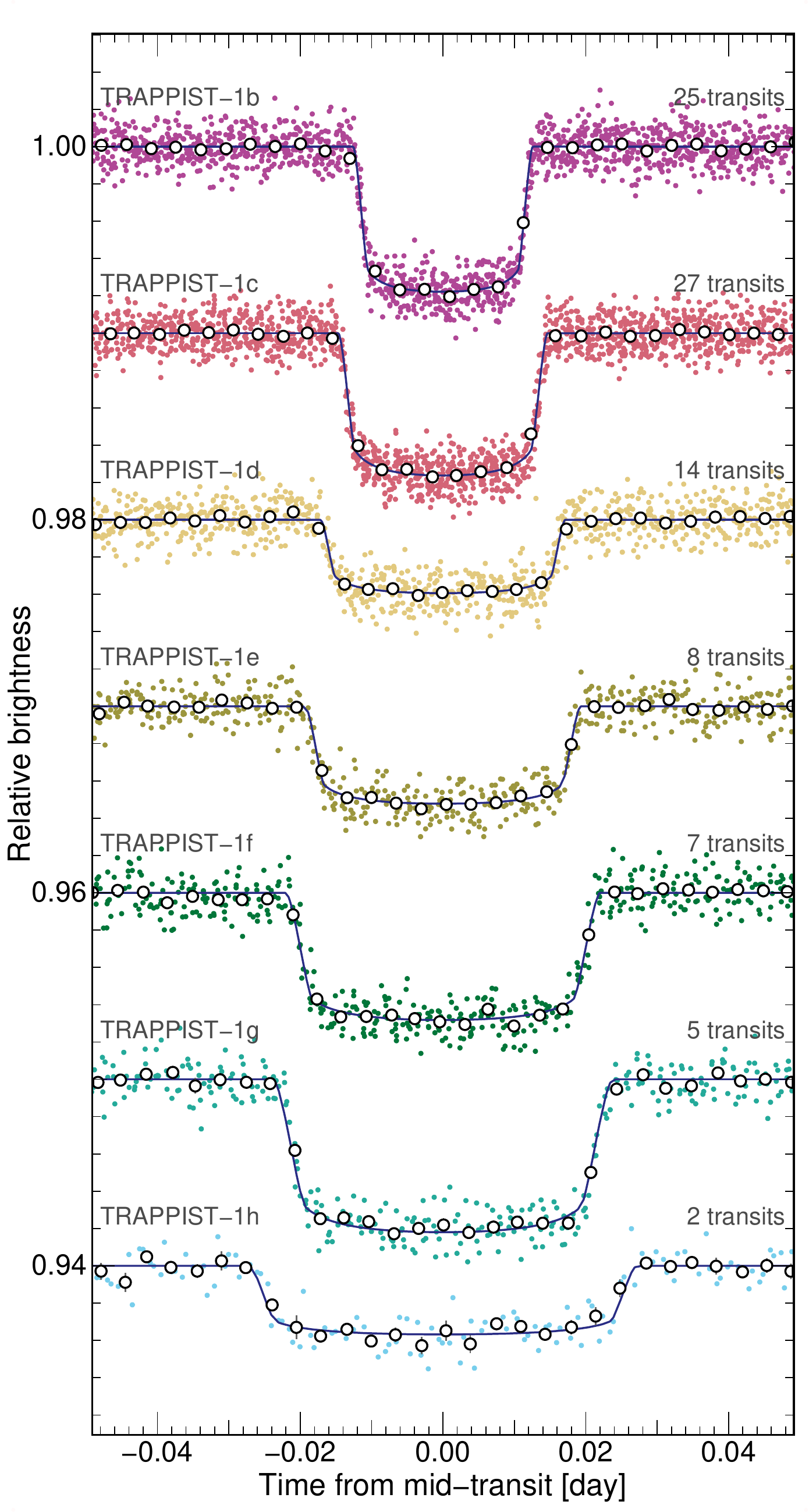}
\end{subfigure}
\begin{subfigure}[b]{0.483\textwidth}
	\includegraphics[width= \textwidth]{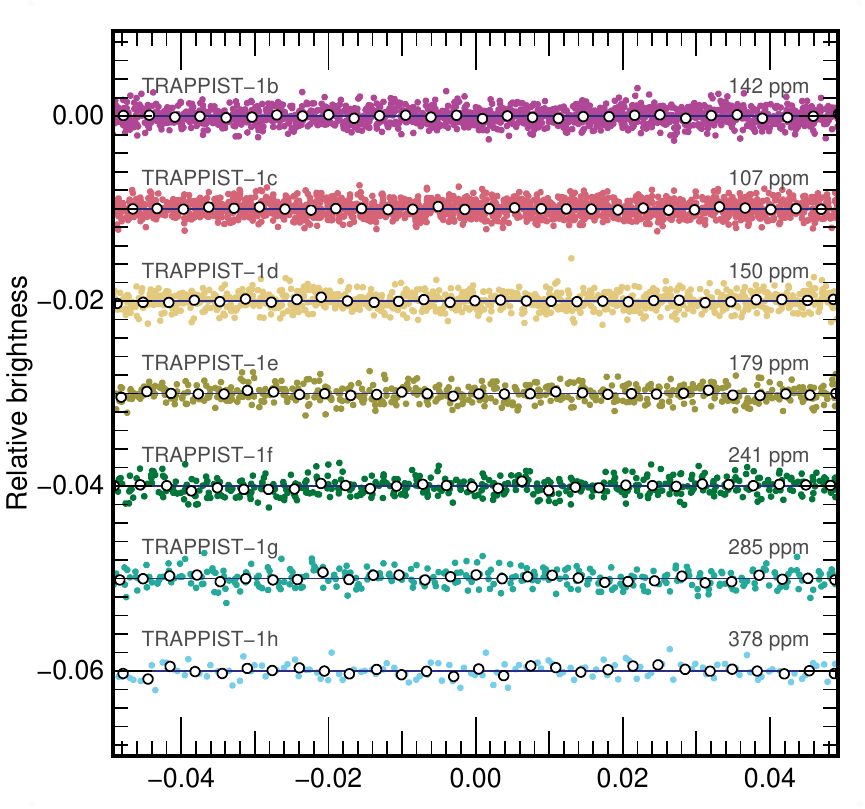}
	\includegraphics[width= 0.98\textwidth]{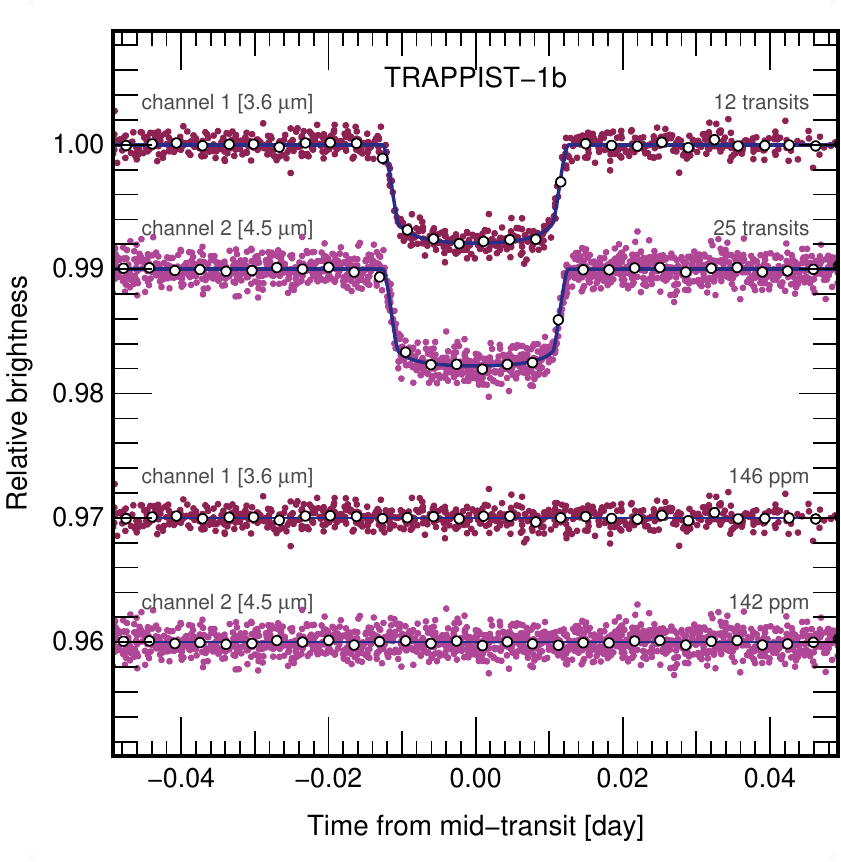}
\end{subfigure}
\caption{\textit{Left:} Period-folded photometric measurements obtained by \textit{Spitzer} at 4.5 $\mu$m near the transits of planets TRAPPIST-1b to \hbox{TRAPPIST-1h}, corrected for the measured TTVs. Coloured dots show the unbinned measurements; open circles depict 5min-binned measurements for visual clarity. The best-fit transit models are shown as coloured lines. The number of transits that were observed to produce these combined curves is written on the plot.
\textit{Top right:} Corresponding residuals. The RMS of the residuals (5min bins) is indicated over each planet. 
\textit{Bottom right:} Similar to other panels, only for TRAPPIST-1b at 3.6 $\mu$m (channel 1) and \hbox{4.5 $\mu$m} (channel 2).
}\label{fig:lc}  
\vspace{0.3cm}
\end{figure*}

In a second phase, we carried out a global MCMC analysis of all the TRAPPIST-1 transits observed by \textit{Spitzer} to improve the determination of the system parameters. We first performed a preliminary analysis, composed of one chain of \hbox{10 000} steps, to determine the correction factors $CF$ to be applied to the error bars of each light curve (see \citealt{Gillon2012} and Appendix~\ref{plotrms} for details). With the corrected error bars, we then performed the final global analysis, consisting of two Markov chains of \hbox{100 000} steps. The jump parameters in our global analysis were as follows:
\begin{itemize}
    \item The stellar mass $M_{\star}$, radius $R_{\star}$, effective temperature $T_{\rm{eff}}$, and metallicity [Fe/H].
    \item The linear combinations $c_1$ and $c_2$ of the quadratic limb-darkening coefficients ($u_{1}$, $u_{2}$) for each bandpass, defined as previously.
    \item For the seven planets, the transit depth d$F_{\mathrm{4.5\mu m}}$ at \hbox{4.5 $\mu$m}, and the transit impact parameter $b$. The transit duration was not a jump parameter anymore in the global analysis, as it is uniquely defined for each planet by its orbital period, transit depth, and impact parameter, combined with the stellar mass and radius \citep[][]{Seager2003}.  This assumption neglects orbital eccentricity and transit duration variations, which may be justified due to the small eccentricities expected when migrating into Laplace resonances \citep{Luger2017}, and the small amplitude of transit duration variations that is expected based on dynamical models.
    \item For TRAPPIST-1b, the transit depth difference between the \textit{Spitzer}/IRAC 3.6-$\mu$m and 4.5-$\mu$m channels: \hbox{$\mathrm{dd}F=\mathrm{d}F_{\mathrm{3.6\mu m}}-\mathrm{d}F_{\mathrm{4.5\mu m}}$.}
    \item For the six inner planets, the transit timing variation (TTV) of each transit with respect to the mean transit ephemeris derived from the individual analyses (cf. Section \ref{indiv_analyses}).
    \item For TRAPPIST-1h, the orbital period $P$ and the mid-transit time $T_0$.
\end{itemize}
This gives a total of 122 jump parameters for 19 258 data points. As previously, we assumed circular orbits for all the planets. For $M_{\star}$, we used a normal prior PDF based on the mass of \hbox{0.089 $\pm$ 0.007 $M_{\odot}$} semi-empirically derived by \cite{VanGrootel2017} for TRAPPIST-1 by combining a prior from stellar evolution models to a set of dynamical masses recently reported by \cite{Dupuy2017} for a sample of equivalently classified ultracool dwarfs in astrometric binaries. We prefer here to assume this semi-empirical prior for the stellar mass rather than a purely theoretical one, like the one used previously in \citetalias{Gillon2016} and \citetalias{Gillon2017}, as current stellar evolutionary models are known to underestimate the radii of some low-mass stars (e.g. \citealt{Torres2013}, \citealt{MacDonald2014}, and references therein; see \citealt{VanGrootel2017} for more details). We assumed the same normal prior PDFs as previously for [Fe/H], and ($u_1$, $u_2$) for both bandpasses. Uniform non-informative prior distributions were assumed for the other jump parameters.\\
\indent
The convergence of the chains was again checked with the statistical test of \cite{Gelman1992}. The Gelman-Rubin statistic was less than 1.11 for every jump parameter, measured across the two chains, indicating that the chains are converged. We also estimated the effective sample size of the chains ($N_{\mathrm{eff}}$) by computing the integrated autocorrelation length as defined in \cite{Gelman2013}. We find a minimum $N_{\mathrm{eff}}$ of 27, with a median value of 118 over all parameters. Fig.~\ref{fig:acf} shows the autocorrelation function versus time lag for all 122 jump parameters.\\
\indent
The physical parameters of the system were deduced from the jump parameters at each step of the MCMC, so that their posterior PDFs could also be constructed. At each MCMC step, the value for $R_{\star}$ was combined with the updated luminosity reported by \cite{VanGrootel2017}, \hbox{$L_{\star}$ = (5.22 $\pm$ 0.19) $\times$ 10$^{-4}$ $L_{\odot}$}, based on their improved measurement of the star's parallax, to derive a value for $T_{\mathrm{eff}}$. For each planet, values for $R_{\mathrm{p}}$, $a$, and $i$, were deduced from the values for the stellar and transit parameters. Finally, values were also computed for the irradiation of each planet in Earth units and for their equilibrium temperatures, assuming a null Bond albedo.\\
\indent
Fig. \ref{fig:lc} shows the detrended period-folded photometry for each planet with the corresponding best-fit transit model, while Figs.~\ref{fig:RMS1}, \ref{fig:RMS2}, \ref{fig:RMS3}, \ref{fig:RMS4}, and \ref{fig:RMS5} display binned residuals RMS vs. bin size plots for the 78 light curves of our dataset.
Figs. \ref{fig:global_correlations1} and \ref{fig:global_correlations3} show the cross-correlation plots and histograms of the posterior PDFs derived for the jump parameters of our global analysis, while Table \ref{table_glob} presents the parameters derived for the system. We discuss these results in Section \ref{discussion}.

\begin{table*}
\centering
\caption{Updated system parameters: median values and 1-$\sigma$ limits of the posterior PDFs derived from our global MCMC analysis.}
\begin{scriptsize}
\begin{tabular}{lccccccc}
\hline
\textbf{Parameters} & \multicolumn{7}{c}{\textbf{Value}} \\
\hline
\vspace{0.12cm}
\bf{Star} & \multicolumn{7}{c}{\textbf{TRAPPIST-1}} \\
\vspace{0.12cm}
Mass$^{a}$, $M_{\star}$ ($M_{\odot}$) & \multicolumn{7}{c}{$0.0890 \pm 0.0070$} \\
\vspace{0.12cm}
Radius, $R_{\star}$ ($R_{\odot}$) & \multicolumn{7}{c}{$0.1210 \pm 0.0030$} \\
\vspace{0.12cm}
Density, $\rho_{\star}$ ($\rho_{\odot}$) & \multicolumn{7}{c}{$51.1_{-2.4}^{+1.2}$} \\
\vspace{0.12cm}
Luminosity$^{a}$, $L_{\star}$ ($L_{\odot}$) & \multicolumn{7}{c}{${0.000522 \pm 0.000019}$} \\
Effective temperature, & \multicolumn{7}{c}{$2511 \pm 37$} \\
\vspace{0.12cm}
$T_{\mathrm{eff}}$ (K) & & & & & & & \\
\vspace{0.1cm}
Metallicity$^{a}$, [Fe/H] (dex) & \multicolumn{7}{c}{$+0.04 \pm 0.08$} \\
\vspace{0.1cm}
Limb-darkening coefficient$^{a}$, $u_{1,\:\mathrm{4.5\mu m}}$ & \multicolumn{7}{c}{$0.161 \pm 0.019$} \\
\vspace{0.1cm}
Limb-darkening coefficient$^{a}$, $u_{2,\:\mathrm{4.5\mu m}}$ & \multicolumn{7}{c}{$0.208 \pm 0.021$} \\
\vspace{0.1cm}
Limb-darkening coefficient$^{a}$, $u_{1,\:\mathrm{3.6\mu m}}$ & \multicolumn{7}{c}{$0.168 \pm 0.020$} \\
\vspace{0.1cm}
Limb-darkening coefficient$^{a}$, $u_{2,\:\mathrm{3.6\mu m}}$ & \multicolumn{7}{c}{$0.244 \pm 0.021$} \\
\hline
\vspace{0.12cm}
\bf{Planets} & \bf{b} & \bf{c} & \bf{d} & \bf{e} & \bf{f} & \bf{g} & \bf{h} \\
Number of & 37 & 27 & 14 & 8 & 7 & 5 & 2 \\
\vspace{0.12cm}
transits observed & & & & & & & \\
\vspace{0.12cm}
Period, $P$ (days) & 
1.51087637 &
2.42180746 &
4.049959 &
6.099043 & 
9.205585 & 
12.354473 & 
18.767953 \\
\vspace{0.12cm}
& $ \pm 0.00000039$ &
 $ \pm 0.00000091$ &
 $ \pm 0.000078$ &
 $ \pm 0.000015$ &
 $ \pm 0.000016$ &
 $ \pm 0.000018$ &
 $ \pm 0.000080$ \\
Mid-transit time, & 
7322.51654 &
7282.80879 & 
7670.14227 & 
7660.37910 & 
7671.39470 & 
7665.35084 &
7662.55467 \\
\vspace{0.12cm}
$T_{0}-2,450,000$ ($\mathrm{BJD_{TDB}}$)&
$ \pm 0.00012$ &
 $ \pm 0.00018$ &
 $ \pm 0.00026$ &
 $ \pm 0.00040$ &
 $ \pm 0.00022$ &
 $ \pm 0.00020$ &
 $ \pm 0.00054$ \\
Transit depth at 4.5 $\mu$m, & 
$0.7277 \pm 0.0075$ & 
$0.6940 \pm 0.0068$ & 
$0.3566 \pm 0.0070$ & 
$0.4802 \pm 0.0094$ & 
$0.634 \pm 0.010$ & 
$0.764 \pm 0.011$ &
$0.346 \pm 0.014$ \\
\vspace{0.12cm}
d$F_{\mathrm{4.5\mu m}}$ ($\%$) & & & & & & & \\
Transit depth at 3.6 $\mu$m & $0.7070 \pm 0.0086$ & - & - & - & - & - & - \\
\vspace{0.12cm}
d$F_{\mathrm{3.6\mu m}}$ ($\%$) & & & & & & & \\
Transit impact & $0.157 \pm 0.075$ &
$0.148 \pm 0.088$ &
$0.08_{-0.06}^{+0.10}$ & $0.240_{-0.047}^{+0.056}$ & $0.337_{-0.029}^{+0.040}$ & $0.406_{-0.025}^{+0.031}$ & $0.392_{-0.043}^{+0.039}$ \\
\vspace{0.12cm}
parameter, $b$ ($R_{\star}$) & & & & & & & \\
\vspace{0.12cm}
Transit duration, $W$ (min) & 
$36.19 \pm 0.12$ & 
$42.31 \pm 0.14$ & 
$49.33_{-0.32}^{+0.43}$ & 
$55.92 \pm 0.39$ & 
$63.14 \pm 0.36$ & 
$68.53 \pm 0.37$ & 
$76.92 \pm 0.96$ \\
\vspace{0.12cm}
Inclination, $i$ ($^{\circ}$) & 
$89.56 \pm 0.23$ & 
$89.70 \pm 0.18$ & 
$89.89_{-0.15}^{+0.08}$ & $89.736_{-0.066}^{+0.053}$ & $89.719_{-0.039}^{+0.026}$ & $89.721_{-0.026}^{+0.019}$ & 
$89.796 \pm 0.023$ \\
Semi-major axis, & $11.50_{-0.25}^{+0.28}$ & $15.76_{-0.34}^{+0.38}$ & $22.19_{-0.48}^{+0.53}$ & $29.16_{-0.63}^{+0.70}$ & $38.36_{-0.84}^{+0.92}$ & 
$46.7 \pm 1.1 $ & $61.7_{-1.3}^{+1.5}$ \\
\vspace{0.12cm}
$a$ ($10^{-3}$ AU) & & & & & & & \\
\vspace{0.12cm}
Scale parameter, $a/R_{\star}$ & $20.56_{-0.31}^{+0.16}$ & $28.16_{-0.44}^{+0.22}$ & 
$39.68_{-0.62}^{+0.32}$ & 
$52.13_{-0.82}^{+0.41}$ & 
$68.6_{-1.1}^{+0.6}$ & 
$83.5_{-1.3}^{+0.7}$ & $110.3_{-1.7}^{+0.9}$ \\
\vspace{0.12cm}
Irradiation, $S_{\mathrm{p}}$ ($S_{\oplus}$) & $3.88 \pm 0.22$ & 
$2.07 \pm 0.12$ & 
$1.043 \pm 0.060$ & 
$0.604 \pm 0.034$ & 
$0.349 \pm 0.020$ &
$0.236 \pm 0.014$ & 
$0.135_{-0.074}^{+0.078}$\\
Equilibrium & & & & & & & \\
\vspace{0.12cm}
temperature$^{b}$ $T_{\mathrm{eq}}$ (K) & 
$391.8 \pm 5.5$ & 
$334.8 \pm 4.7$ & 
$282.1 \pm 4.0$ & 
$246.1 \pm 3.5$ & 
$214.5 \pm 3.0$ & 
$194.5 \pm 2.7$ & 
$169.2 \pm 2.4$ \\
\vspace{0.12cm}
Radius, $R_{\mathrm{p}}$ ($R_{\oplus}$) & 
$1.127 \pm 0.028$ & 
$1.100 \pm 0.028$ & 
$0.788 \pm 0.020$ & 
$0.915 \pm 0.025$ & 
$1.052 \pm 0.026$ & 
$1.154 \pm 0.029$ & 
$0.777 \pm 0.025$ \\
\hline
\end{tabular}
\end{scriptsize}
\begin{tablenotes}
\item[] 
\textbf{Notes.} $ ^{a}$ Informative prior PDFs were assumed for these stellar parameters (see Section \ref{global_analysis}).\\
$ ^{b}$Assuming a null Bond albedo.
\end{tablenotes}
\label{table_glob}
\end{table*}

\subsection{Stellar variability at 4.5 $\mu$m} 
\label{overall_variability}

\begin{figure}
\vspace{-0.3cm}
\centering
\includegraphics[width=0.48\textwidth]{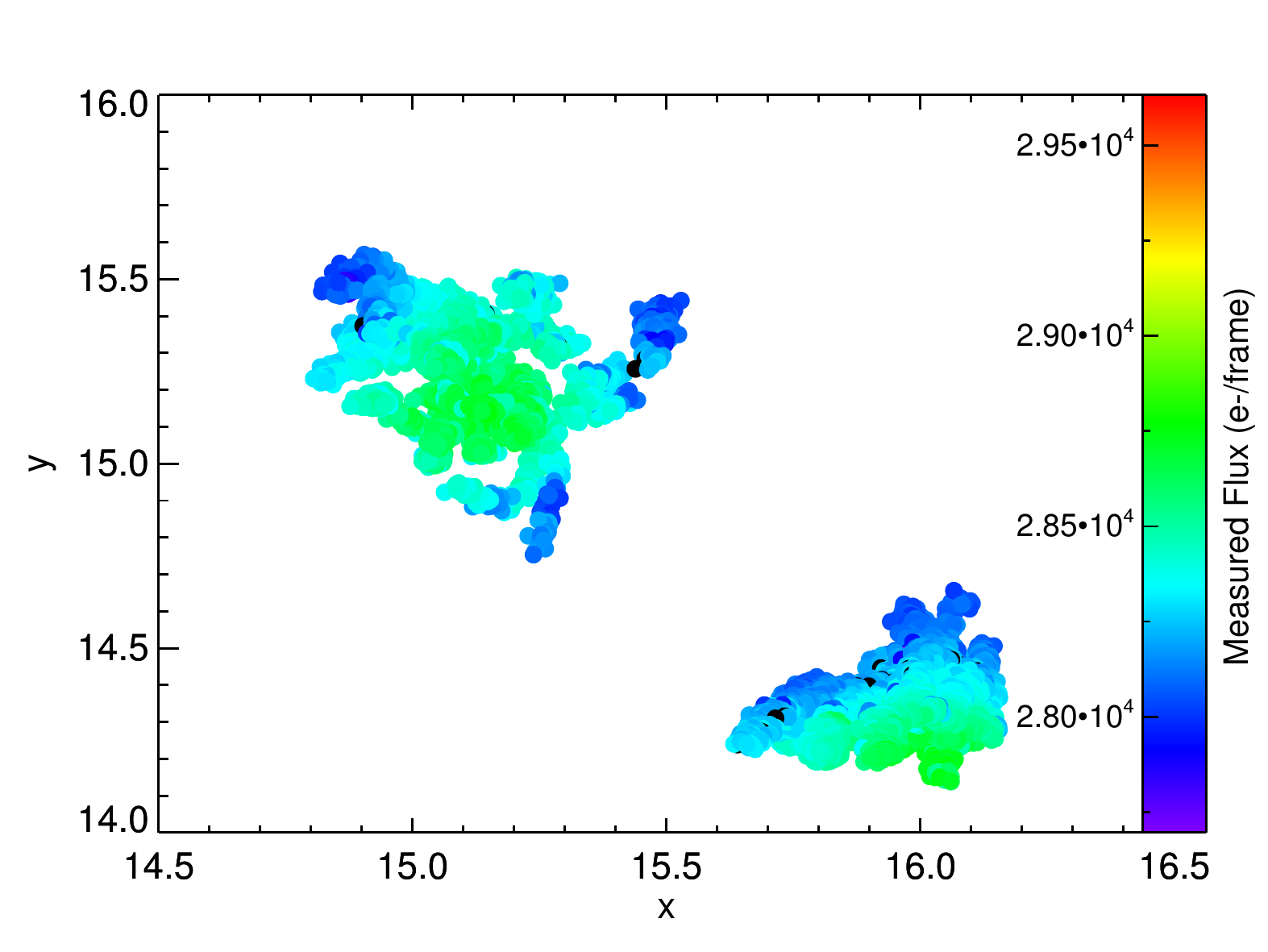}
\vspace{-0.6cm}
\caption{TRAPPIST-1 \textit{Spitzer}/IRAC raw photometric fluxes collected from all 66 AORs taken between 2016 and March 2017 and the corresponding centroid locations on the detector.}\label{fig:fluxes} 
\vspace{-0.5cm}
\end{figure}

\begin{figure}
\vspace{-1.3cm}
\centering
\includegraphics[width=0.53\textwidth]{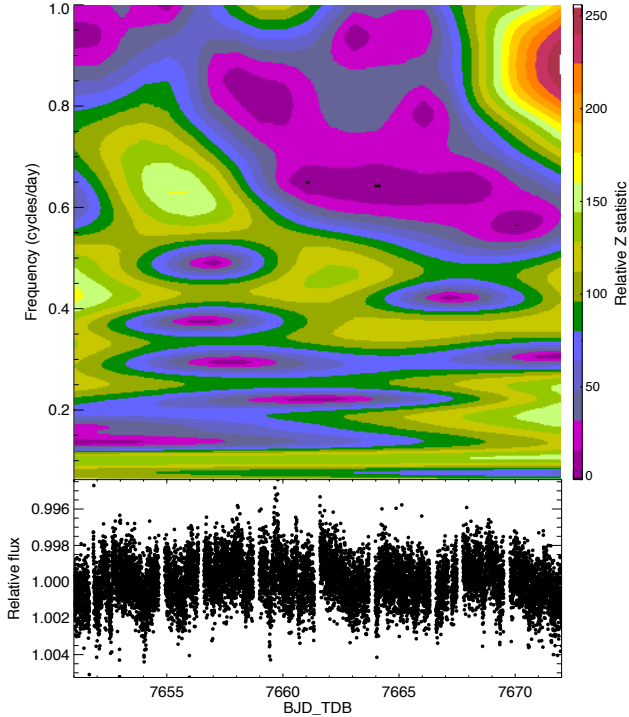}
\vspace{-1.3cm}
\caption{Wavelet diagram (\textit{top}) of the photometric residuals (\textit{bottom}).}\label{fig:wavelet}
\end{figure}

Monitoring the stellar variability with \textit{Spitzer} is rendered difficult because of two factors. First, half of the \textit{Spitzer} observations analysed in this paper have been obtained in a time sparse mode, focusing on the transit windows of the seven TRAPPIST-1 planets. These sequences are short, up to four hours long, which is not enough to sample the rotational period of the star. Second, we failed to consistently position the target on the detector's sweet spot, which systematically affects the measured flux at the $\sim$2\% level. Figure~\ref{fig:fluxes} illustrates the absolute flux level measured for all TRAPPIST-1 AORs included in the present paper and the corresponding centroid locations on the detector. Two distinct areas can be seen because of pointing inaccuracies due to the target's large differential parallax between the Earth and \textit{Spitzer}. To mitigate both caveats, we elect to conduct our variability analysis independently from the global fit presented above by only including the quasi-continuous sequence obtained over 21 days in Sept-Oct 2016 (\citetalias{Gillon2017}). The corresponding centroid locations are clustered on bottom right of Figure~\ref{fig:fluxes}.\\ 
\indent
We performed the data reduction by computing the absolute fluxes of all AORs using a fixed aperture size of 3 pixels throughout the dataset. A complication arose from the removal of the pixel-phase effect as TRAPPIST-1 fell 1 to 2 pixels away from the detector's sweet spot for which a high-resolution gain map exists \citep{Ingalls:2016}. As no such map was available for our purpose, we calibrated the absolute photometry by using the data itself, using an implementation of the BLISS mapping algorithm \citep{Stevenson2012}. We found that the entire area over which the star fell is relatively extended (0.6 pixels along the $x$-axis and 0.5 pixels along the $y$-axis), which marginally limits the flux calibration accuracy. For the purposes of this stellar variability analysis, we discarded flares and removed transits based on the parameters deduced from our global analysis (Table~\ref{table_glob}). The photometric residuals are thus assumed to include signal from the star alone. We measured a photometric residual RMS of 0.11\% at a 123-s cadence. We performed a discrete Fourier analysis of the residuals that yielded a maximum at $\sim$10 days. The $3.30 \pm 0.14$-day rotation period found by \citeauthor{Luger2017} (\citeyear{Luger2017}, see also \citealt{2017ApJ...841..124V}) using 80-day continuous observations of visible K2 data, appears as a low-amplitude peak in our periodogram.\\ 
\indent
To obtain a more detailed view of the signal components, we further perform a wavelet analysis of the photometric residuals. For this purpose, we use the weighted wavelet Z-transform (WWZ) code presented in \cite{Foster1996}. The wavelet Z-statistic is computed as a function of both time and frequency, which gives further insights into the structure of the photometric residuals. The results of this analysis are shown on Figure~\ref{fig:wavelet}. We find that multiple peaks exist but that are of low amplitude. No signal is apparent around 3.3 days but the $\sim$10-day signature seems to persist across the entire window, albeit of low significance. The low-amplitude residual correlated noise could originate from imperfect instrumental systematic correction or from stellar noise. Assuming that the systematics correction we use is efficient, the wavelet analysis applied to the photometric residuals suggests that the star exhibits multiple active regions that evolve rapidly with time. The data at hand does not enable us to clearly identify the structure of the signal. We argue that long-term parallel monitoring of TRAPPIST-1 in the visible and infrared are desirable to better constrain its variability patterns.

\indent
We carried out an additional analysis of the power spectrum of TRAPPIST-1 using a correlated noise model described by a Gaussian process with a power spectrum defined as the sum of stochastically-driven damped simple-harmonic oscillators, each with quality factor $Q$ and frequency $\omega_0$ \citep{ForemanMackey2017}. We fit the data with two components: a fixed $Q=1/\sqrt{2}$ term which has a power spectrum describing variability similar to granulation, and a second $Q >>1$ term describing (quasi-)periodic variability, with a frequency initialized for a period of $3$ days. We found that the amplitude of the large-$Q$ term decreased to zero when optimizing the parameters of the Gaussian process, indicating that there is no evidence for quasi-periodic variability in the \textit{Spitzer} \hbox{4.5-$\mu$m} dataset. We found that the ``granulation" term had a finite amplitude with a variance of $7\times 10^{-4}$, and a frequency $\omega_0 = 22.45$ rad day$^{-1}$, corresponding to a characteristic damping timescale of 0.28 days. Once the power spectrum was optimized, we subtracted off the Gaussian process estimate of the correlated noise component, and found that the normalized residuals follow a Gaussian, but slightly broader by a factor of 1.065, which is an additional argument for increasing the uncertainties on the data points with the correction factors, $CF$, as discussed above.


\section{Discussion}
\label{discussion}

\begin{figure*}
\center
\begin{subfigure}[b]{0.49\textwidth}
	\includegraphics[width= \textwidth]{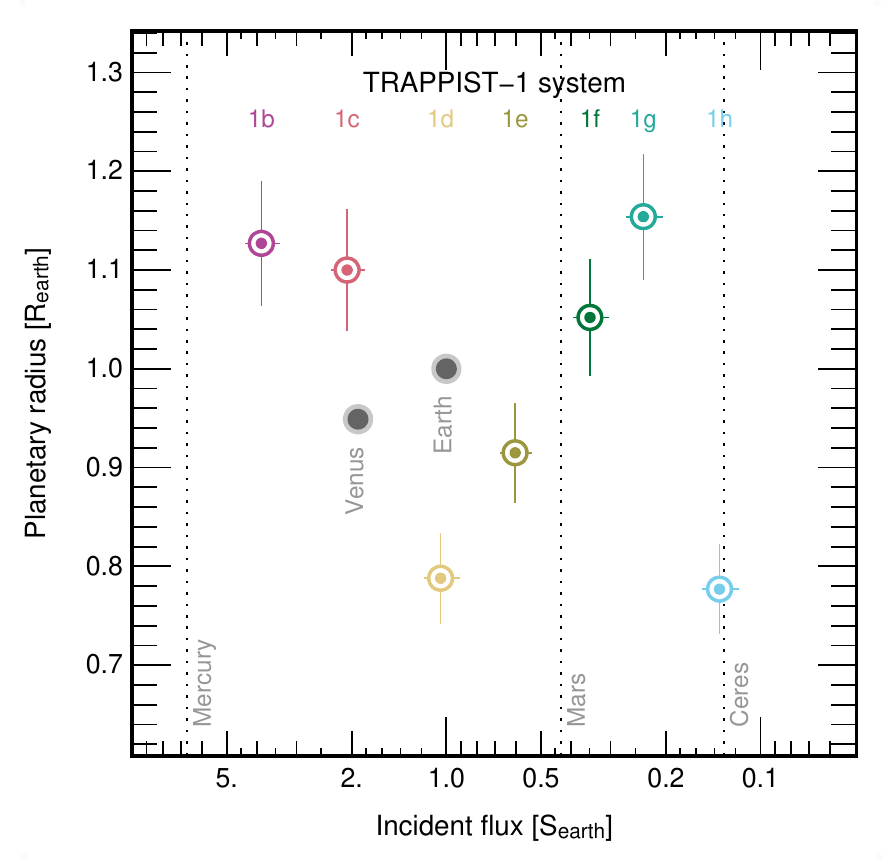}
\end{subfigure}
\begin{subfigure}[b]{0.49\textwidth}
	\includegraphics[width= \textwidth]{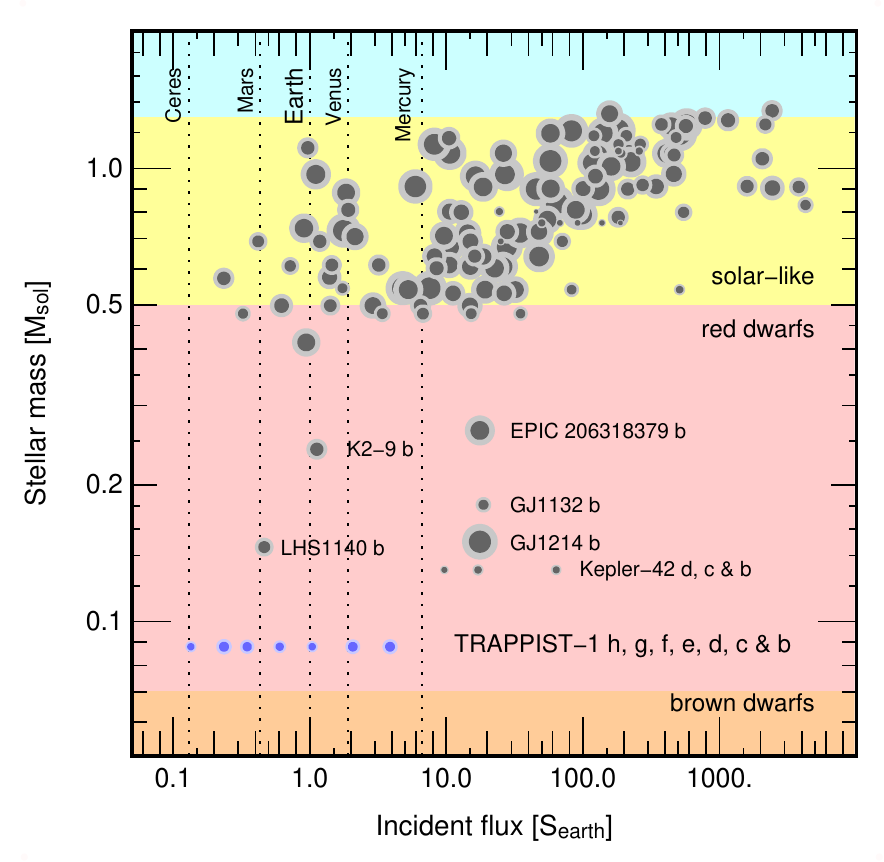}
\end{subfigure}
\caption{Updates on diagrams shown in \citetalias{Gillon2016} and \citetalias{Gillon2017}. \textit{Left:} Planetary radii as a function of incident flux.
\textit{Right:} Planet population shown ordered by stellar-host mass as a function of incident flux. Only planets with radii $<2 R_\oplus$ are represented. The dots size increases linearly with radius.
}\label{fig:pop}
\vspace{0.5cm}
\end{figure*}

\subsection{Updated system parameters}

We have more than doubled the number of transit events observed with \textit{Spitzer} on TRAPPIST-1 with respect to what has been presented in \citetalias{Gillon2017}. Our global analysis refines the planets' transit depths (at 4.5 $\mu$m) by factors up to 2.8 (TRAPPIST-1e) and their transit durations by factors up to 2.8 (TRAPPIST-1h), and also slightly improves the precision of the physical parameters derived for the planets. An important point to outline is that in \citetalias{Gillon2016} and \citetalias{Gillon2017} our global analysis assumed an informative prior on the stellar mass and radius - and thus on the stellar density - based on stellar evolution models. In this new analysis, no informative prior was assumed for the stellar radius, and the stellar density was only constrained by the shape of the transits of the seven planets \citep[][]{Seager2003}. Furthermore, our assumed prior on the stellar mass was here not only based on stellar physics computations but also on empirical data (see 
Section \ref{global_analysis} and \citealt[][]{VanGrootel2017}). We thus consider our updated planetary parameters presented in Table \ref{table_glob} not only more precise than those reported in \citetalias{Gillon2017}, but also more accurate because they are less model-dependent. We compare some of these quantities to Solar system objects and the rest of the small exoplanet population in Fig.~\ref{fig:pop}.

\subsection{Variability of the transit parameters} 
\label{sec:variability}

\begin{figure}
\centering
\includegraphics[width=0.47\textwidth]{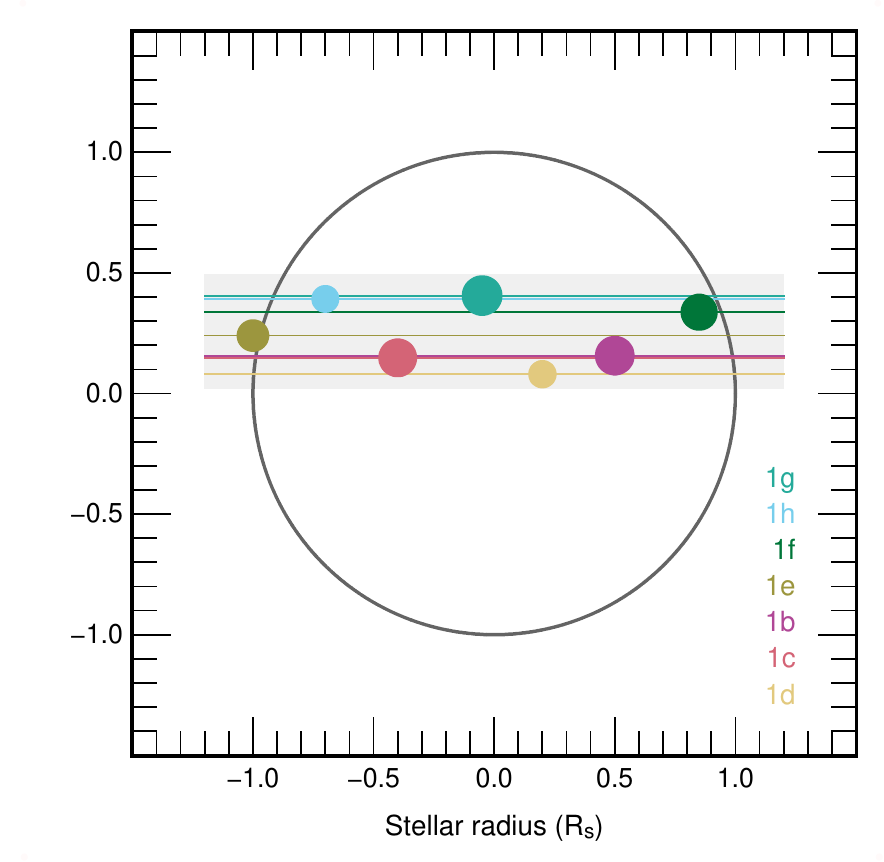}
\caption{Representation of the star, and of the planets that transit it, using impact parameters and transit depths from Table~\ref{table_glob}. The planets cover a minimum area of $\sim$56$\%$ of an hemisphere. Planetary position on its chord is arbitrary.
}\label{fig:chord} 
\vspace{0.5cm}
\end{figure}

As reported in Section \ref{overall_variability}, we do not find any
significant (quasi-)periodic signal in the IRAC \hbox{4.5-$\mu$m} band related to the rotation of TRAPPIST-1. Rotational variability has however been previously detected by \cite{Luger2017} in the {\it K2} passband, at the level of few percents and with a period of $\sim$3.3 days. This periodic photometric variability indicates that inhomogeneities of the stellar surface (spots) move in and out of view as the star rotates. \cite{Luger2017occ} recently demonstrated that the orbital planes of the TRAPPIST-1 planets are aligned to $<$0.3$^{\circ}$ at 90\% confidence. Together, the planets cover at least 56\% of a hemisphere when they transit (shadowed area in Fig.~\ref{fig:chord}), notably the low and intermediate latitudes at which we find spots on the Sun (e.g. \citealt{Miletskii2009}). We could therefore expect the planets to cross some stellar spots during their transits. Such occulted spots would affect the transit profile in a way that would depend on their size, contrast, and distribution across the planetary chord, and would tend to make the transit shallower. Unocculted spots, i.e., stellar spots that are not crossed during planetary transits, would have the opposite effect, making the transit deeper by diminishing the overall flux from the star while leaving the surface brightness along the transit chord unchanged. Because of stellar spots, either occulted or unocculted, we could thus expect to detect variations in the planets' transit depths as a function of time.\\
\indent
As noted in Section~\ref{sec:indi}, the transit depths derived from the individual analyses generally follow a normal distribution for each planet, except TRAPPIST-1b at 3.6 $\mu$m, TRAPPIST-1e, and TRAPPIST-1g, each of which show some outliers. It is common knowledge that transit parameters derived from a single transit light curve can be significantly affected by systematics (e.g. \citealt{Gillon2012}). The global analysis of an extensive set of transit light curves, which assumes one unique transit profile for all the transit light curves (of a same planet), allows a better separation of the actual transit signal from the correlated noise and the derivation of robust transit parameters. To better assess the possible variations in the transit depth of each planet, we computed for each of its transits the median values of the photometric residuals in transit and out of transit, using for this purpose the photometric residuals from the global analysis. These median values, together with the median absolute deviations, are given for each transit in Table \ref{tab:discuss}. A significant difference between the in-transit and out-of-transit medians of the photometric residuals for a given transit would indicate a variation in its depth compared to the planet's transit depth derived from the global analysis. However, the in-transit and out-of-transit medians are compatible within 1 sigma for all transits of all planets. This test reveals that the variability seen for some planets (TRAPPIST-1b at 3.6 $\mu$m, TRAPPIST-1e, and TRAPPIST-1g) in the transit depths derived from the individual analyses is likely not physical but rather caused by systematic effects --which are better disentangled from the planetary signals in the global analysis.\\ 
\indent
In this context, it is worth noting that instrumental systematics are stronger at 3.6 $\mu$m than at \hbox{4.5 $\mu$m} and thus require more complex baseline models, which can introduce some biases in the transit parameters derived from a single light curve. This might explain the increased scatter of TRAPPIST-1b's individual transit depths at 3.6 $\mu$m. For TRAPPIST-1e, we note that only the first transit in Fig.~\ref{fig:singles} (epoch -1) shows a discrepant transit depth. This transit was actually observed over two different consecutive AORs, which might have introduced a bias in its measured depth. As for \hbox{TRAPPIST-1g}, a visual inspection of the global analysis' residuals for the two transits with discrepant transit depths (second and fifth transits in Fig.~\ref{fig:singles}, corresponding to epochs 0 and 14 respectively) does not reveal any obvious structures or spot-crossings. The origin of these two outliers thus remains unclear. Additional transit observations of this planet at higher SNR are needed to better assess the possible variability of its transit depth.\\
\indent
Overall, {\it Spitzer} transits of TRAPPIST-1 planets thus appear to be mostly immune to the effects of stellar variability. There are several reasons why this may happen:
\begin{itemize}
\item Considering the very low level of low-frequency photometric variability shown by TRAPPIST-1 at 4.5 $\mu$m, unocculted spots may not have a significant impact on the planets' transit depths. Using the simple model of \citeauthor{Berta2011} (\citeyear{Berta2011}, see their Equations 8 and 9) to estimate the expected amplitude of transit depth variations at 4.5 $\mu$m due to unocculted spots based on the stellar variability measured in that band, we indeed find amplitudes lower than 100 ppm for all the planets. This is smaller than the error bars on the individual transit depths. We note however that this estimate is only a lower limit to the possible amplitude of transit depth variations due to unocculted spots. Indeed, the rotational variability of the star reflects only the non-axisymmetric component of the stellar surface inhomogeneities. The axisymmetric component does not contribute to the measured stellar variability and its effect is thus not included in our estimate, while it is also expected to affect the planets' transit depths.
\item The periodic variability detected in the {\it K2} passband may be caused by high-latitude spots that do not cross the planets' chords, explaining the non-detection of spot-crossing events. Evidence for such high-latitude spots has been reported for some mid- and late-M dwarfs (see e.g. \citealt{Barnes2015}). We note however that these objects are usually young \hbox{($\lesssim$1 Gyr)}, while TRAPPIST-1 is a rather old system (\hbox{7.6 $\pm$ 2.2 Gyr}, \citealt{Burgasser2017}).
\item Spot-crossing events may not produce detectable effects on the \textit{Spitzer} transit light curves due to the reduced spot-to-photosphere contrast in the IRAC passbands (see e.g. \citealt{Ballerini2012}). A practical example of this effect was presented by \cite{Fraine2014}, who reported simultaneous \textit{Kepler} and \textit{Spitzer} transit photometry of the Neptune-sized planet HAT-P-11b orbiting an active K4 dwarf. While some spot-crossing events are clearly visible in the \textit{Kepler} optical photometry, they are undetected in the photometry obtained concurrently with \textit{Spitzer} at 4.5 $\mu$m.
\end{itemize}

\subsection{Transmission spectrum of TRAPPIST-1b} 

A transmission spectrum can be severely affected by both occulted and unocculted stellar spots \citep[e.g.][]{Jordan2013,McCullough:2014lr,Rackham2017}. To extract a proper transmission spectrum, a planet needs to cover a part of the stellar surface that has a spectrum that is representative of the whole disc. We are in a privileged position with TRAPPIST-1 as the planets cross over a quarter of the entire stellar surface, and more than half of a hemisphere. Unless spots never intersect transit chords, the planets will cross over a representative fraction of the star, and the transmission spectra measured for the TRAPPIST-1 planets should be robust measures of their atmospheres \citep{de-Wit2016,Barstow2016,Morley2017}.\\
\indent
Our global data analysis yields a marginal transit depth difference of $+208 \pm 110$ ppm for TRAPPIST-1b between the \textit{Spitzer}/IRAC 4.5-$\mu$m and 3.6-$\mu$m channels, the transit being slightly deeper at 4.5 $\mu$m than at 3.6 $\mu$m (Table~\ref{table_glob}). If confirmed, this transit depth variation would imply that \hbox{TRAPPIST-1b's} atmosphere significantly exceeds its equilibrium temperature (392~K assuming a null Bond albedo, Table~\ref{table_glob}). A deeper transit depth at 4.5 $\mu$m would be best explained by atmospheric CO$_2$, which has a prominent absorption feature around 4.2 $\mu$m \citep[see e.g.,][]{Kaltenegger2009}. The most favorable scenario to enhance such a signature requires no other opacity source across the IRAC 3.6-$\mu$m channel than the extended wing of the \hbox{4.2-$\mu$m} absorption band of CO$_2$. In such a case, the variation in transit depth is approximately equivalent to 5 scale heights ($H$) at medium spectral resolution \citep[see e.g.,][]{Kaltenegger2009,deWit2013} and 2$H$ when binned over IRAC's channels. \hbox{TRAPPIST-1b's} scale height would thus be larger or equal to approximately 52 km (\hbox{208 ppm =} $((R_\mathrm{p}+2H)/R_\star)^2-(R_\mathrm{p}/R_\star)^2$). For comparison, the Earth's atmospheric scale height is 8.5~km, while that of Venus is 15.9~km\footnote{\url{https://nssdc.gsfc.nasa.gov/planetary/factsheet/}}. This lower limit on the planet atmospheric scale height can be translated into a lower limit on its atmospheric temperature owing to assumptions of its atmospheric mean molecular mass, $\mu$, and surface gravity, $g$. \hbox{Planet b's} surface gravity is approximately 8 m/s$^2$ \citep{Grimm2017}. Under the present assumption that no strong absorber affects the \hbox{3.6-$\mu$m} band, the background gas cannot be water or methane--both exhibit absorption features in this band--implying that the atmosphere must have a large $\mu$ (e.g., \hbox{28 u.m.a.} if dominated by N$_2$). Under such a favorable scenario for the CO$_2$ feature, TRAPPIST-1 b's atmosphere would require an average temperature above 1400~K--more than three times larger than its estimated equilibrium temperature. On the other hand, if water dominates its atmosphere, this would lead to two counter-balancing effects: \hbox{(1) a} decrease of the mean molecular weight and (2) a transit depth variation between the two IRAC channels of no more than a scale height due to water absorption at \hbox{3.3 $\mu$m}. This would yield an atmospheric scale height larger than $\sim$100~km and an average atmospheric temperature above $\sim$1800~K. Other opacity sources, such as clouds or hazes, would require an even larger scale height and atmospheric temperature to support such a variation in transit depth between the two IRAC channels. Therefore, if confirmed, such hint of variability would be indicative of a surprisingly large atmospheric temperature for TRAPPIST-1b.\\
\indent
As demonstrated recently by \cite{Deming2017}, transit measurements of a planet transiting an M-dwarf can be affected by a resolution-linked bias (RLB) effect, which acts to decrease the apparent amplitude of an absorption feature from the planetary atmosphere. This is due to the complex line structure exhibited by M-dwarfs which creates a flux-leakage effect at low to medium spectral resolution wherein the stellar flux does not entirely cancel out in the ratio of in- to out-of-transit flux. We estimated the amplitude of the RLB effect on the transit depth difference expected for TRAPPIST-1b between the IRAC 4.5-$\mu$m and 3.6-$\mu$m channels assuming different atmospheric scenarios, namely H$_2$-, H$_2$O-, N$_2$-, and CO$_2$-rich atmospheres (\citealt{de-Wit2016}). For TRAPPIST-1 spectrum, we used the PHOENIX/BT-Settl model with $T_{\mathrm{eff}}$/log $g$/[M/H] of 2500K/5.5/0.0 (\citealt{Allard2012}). We find a maximal RLB effect of $\sim$115 ppm over a $\sim$2400 ppm absorption feature of methane in a H$_2$-rich atmosphere, which corresponds to a dampening of the methane absorption feature by $\sim$5\%. Although the amplitude of the effect in this case is comparable to the uncertainty on our measured transit depth difference, it is not relevant for the interpretation of our measurements as the hint of feature detected between the two IRAC channels is not of the order of $\sim$2400 ppm. We find amplitudes ranging from only a few ppm to 30 ppm for the other atmospheric scenarios. The marginal difference in transit depth that we measure between both IRAC bands is thus not expected to be significantly affected by the RLB effect.


\section{Conclusions} 
\label{conclusion}

In this work, we presented 60 new transits of the TRAPPIST-1 planets observed with \textit{Spitzer}/IRAC in early 2017. We performed a global analysis of the entire \textit{Spitzer} dataset gathered so far, which enabled us to refine the transit parameters and to provide revised values for the planets' physical parameters, notably their radii, using updated properties for the host star. As part of this study, we also extracted precise transit timings that will be instrumental for TTV studies of the system, to be presented in a companion paper. In addition, we found that the star shows a very low level of low-frequency variability in the IRAC \hbox{4.5-$\mu$m} channel. We did not detect any evidence of a (quasi-)periodic signal related to stellar rotation and found that the planets' transit depths measured with \textit{Spitzer} are mostly not affected by stellar variability. Finally, we also found for TRAPPIST-1b a marginal transit depth difference of $+208\pm110$ ppm between the IRAC 4.5-$\mu$m and 3.6-$\mu$m channels. If confirmed, this transit depth variation could indicate the presence of CO$_2$ in the planet's atmosphere as well as a surprisingly large atmospheric temperature. Together, these results improve our understanding of this remarkable system and help prepare the detailed atmospheric characterization of its planets with JWST.

\section*{Acknowledgements}

We thank E. Gillen for interesting discussions and valuable suggestions. This work is based in part on observations made with the \textit{Spitzer} Space Telescope, which is operated by the Jet Propulsion Laboratory, California Institute of Technology, under a contract with NASA. This work was partially supported by a grant from the Simons Foundation (PI Queloz, grant number 327127). The research leading to these results also received funding from the European Research Council (ERC) under the FP/2007-2013 ERC grant agreement no. 336480, and under the H2020 ERC grant agreement no. 679030; and from an Action de Recherche Concert\'ee (ARC) grant, financed by the Wallonia-Brussels Federation. L.D. acknowledges support from the Gruber Foundation Fellowship. V.V.G. and M.G. are F.R.S.-FNRS Research Associates. E.J. is F.R.S.-FNRS Senior Research Associate. B.-O.D. acknowledges support from the Swiss National Science Foundation in the form of a SNSF Professorship (PP00P2\_163967). E.B. acknowledges funding by the European Research Council through ERC grant SPIRE 647383. A.J.B. acknowledges funding support from the US-UK Fulbright Scholarship program.




\bibliographystyle{mnras}
\bibliography{biblio.bib}



\appendix

\section{Results from the individual analyses}

\onecolumn

\begin{longtable}{c | c c | c c | c c | c c }
\caption{Median values and 1-$\sigma$ limits of the posterior PDFs deduced for the
timings, depths, and durations of the transits from their individual analyses.}\label{table_indiv}\\
\hline
\hline
Planet & Epoch & Channel & $T_0$ & e$T_0$ & d$F$ & ed$F$ & $W$ & e$W$ \\
 & & & \multicolumn{2}{|c|}{[$\rm{BJD_{TDB}} - 2 450 000$]} & \multicolumn{2}{c|}{[\%]} & \multicolumn{2}{c}{[min]} \\
\hline 
\endfirsthead
\multicolumn{9}{c}{\tablename\ \thetable{} -- \textit{continued from previous page}} \\
\hline
Planet & Epoch & Channel & $T_0$ & e$T_0$ & d$F$ & ed$F$ & $W$ & e$W$ \\
 & & & \multicolumn{2}{|c|}{[$\rm{BJD_{TDB}} - 2 450 000$]} & \multicolumn{2}{c|}{[\%]} & \multicolumn{2}{c}{[min]} \\
\hline 
\endhead
\hline \multicolumn{9}{c}{\textit{Continued on next page}} \\ 
\endfoot
\hline \hline
\endlastfoot
b & 78  & 2 & 7440.36514 & 0.00035 & 0.742 & 0.048 & 36.30 & 1.20\\ 
  & 86  & 2 & 7452.45228 & 0.00014 & 0.757 & 0.025 & 35.76 & 0.40\\
  & 93  & 2 & 7463.02847 & 0.00019 & 0.682 & 0.025 & 36.68 & 0.59\\
  & 218 & 2 & 7651.88743 & 0.00022 & 0.774 & 0.040 & 36.36 & 0.69\\
  & 219 & 2 & 7653.39809 & 0.00026 & 0.684 & 0.031 & 36.19 & 0.68\\
  & 220 & 2 & 7654.90908 & 0.00084 & 0.759 & 0.024 & 35.80 & 2.20\\
  & 222 & 2 & 7657.93129 & 0.00020 & 0.735 & 0.030 & 36.26 & 0.54\\
  & 223 & 2 & 7659.44144 & 0.00017 & 0.746 & 0.029 & 36.84 & 0.58\\ 
  & 224 & 2 & 7660.95205 & 0.00033 & 0.675 & 0.041 & 37.07 & 0.95\\ 
  & 225 & 2 & 7662.46358 & 0.00020 & 0.757 & 0.036 & 36.39 & 0.60\\
  & 226 & 2 & 7663.97492 & 0.00070 & 0.776 & 0.045 & 35.90 & 1.90\\ 
  & 227 & 2 & 7665.48509 & 0.00017 & 0.772 & 0.032 & 36.45 & 0.50\\  
  & 228 & 2 & 7666.99567 & 0.00025 & 0.704 & 0.043 & 36.10 & 0.81\\
  & 229 & 2 & 7668.50668 & 0.00030 & 0.728 & 0.041 & 36.54 & 0.81\\
  & 230 & 2 & 7670.01766 & 0.00034 & 0.751 & 0.048 & 36.75 & 0.89\\
  & 231 & 2 & 7671.52876 & 0.00030 & 0.702 & 0.045 & 36.78 & 0.86\\
  & 318 & 2 & 7802.97557 & 0.00016 & 0.751 & 0.027 & 35.73 & 0.65\\
  & 320 & 2 & 7805.99697 & 0.00016 & 0.699 & 0.023 & 36.29 & 0.50\\
  & 321 & 2 & 7807.50731 & 0.00017 & 0.703 & 0.026 & 36.75 & 0.52\\  
  & 322 & 1 & 7809.01822 & 0.00017 & 0.801 & 0.028 & 36.01 & 0.65\\ 
  & 324 & 2 & 7812.04038 & 0.00020 & 0.703 & 0.027 & 36.64 & 0.59\\
  & 325 & 2 & 7813.55121 & 0.00014 & 0.732 & 0.022 & 36.43 & 0.43\\  
  & 326 & 1 & 7815.06275 & 0.00017 & 0.724 & 0.023 & 36.76 & 0.53\\
  & 327 & 1 & 7816.57335 & 0.00011 & 0.663 & 0.021 & 35.36 & 0.35\\
  & 328 & 2 & 7818.08382 & 0.00015 & 0.723 & 0.026 & 36.41 & 0.46\\
  & 329 & 1 & 7819.59478 & 0.00017 & 0.704 & 0.028 & 36.21 & 0.53\\
  & 330 & 1 & 7821.10550 & 0.00020 & 0.737 & 0.032 & 36.33 & 0.60\\
  & 332 & 2 & 7824.12730 & 0.00018 & 0.737 & 0.029 & 36.29 & 0.55\\  
  & 333 & 1 & 7825.63813 & 0.00018 & 0.706 & 0.033 & 36.19 & 0.60\\
  & 334 & 1 & 7827.14995 & 0.00012 & 0.742 & 0.023 & 36.33 & 0.42\\
  & 335 & 1 & 7828.66042 & 0.00024 & 0.727 & 0.031 & 35.70 & 0.69\\
  & 336 & 2 & 7830.17087 & 0.00021 & 0.708 & 0.032 & 36.44 & 0.60\\
  & 338 & 1 & 7833.19257 & 0.00018 & 0.618 & 0.023 & 35.67 & 0.55\\ 
  & 339 & 1 & 7834.70398 & 0.00016 & 0.682 & 0.020 & 37.14 & 0.49\\
  & 340 & 2 & 7836.21440 & 0.00017 & 0.706 & 0.028 & 35.59 & 0.50\\
  & 341 & 1 & 7837.72526 & 0.00014 & 0.702 & 0.023 & 35.74 & 0.56\\
  & 342 & 1 & 7839.23669 & 0.00017 & 0.808 & 0.025 & 35.76 & 0.56\\
\hline  
c & 70  & 2 & 7452.33470 & 0.00015 & 0.698 & 0.023 & 42.28 & 0.48\\
  & 71  & 2 & 7454.75672 & 0.00066 & 0.644 & 0.037 & 42.60 & 2.00\\
  & 152 & 2 & 7650.92395 & 0.00023 & 0.708 & 0.029 & 43.59 & 0.79\\
  & 153 & 2 & 7653.34553 & 0.00024 & 0.668 & 0.026 & 42.62 & 0.69\\
  & 154 & 2 & 7655.76785 & 0.00040 & 0.679 & 0.050 & 43.10 & 1.00\\
  & 155 & 2 & 7658.18963 & 0.00024 & 0.668 & 0.030 & 42.11 & 0.66\\
  & 156 & 2 & 7660.61168 & 0.00051 & 0.681 & 0.056 & 42.20 & 1.30\\
  & 157 & 2 & 7663.03292 & 0.00028 & 0.710 & 0.031 & 42.46 & 0.88\\
  & 158 & 2 & 7665.45519 & 0.00025 & 0.719 & 0.030 & 42.18 & 0.65\\
  & 159 & 2 & 7667.87729 & 0.00031 & 0.700 & 0.038 & 42.54 & 0.81\\
  & 160 & 2 & 7670.29869 & 0.00035 & 0.731 & 0.044 & 42.34 & 0.96\\
  & 215 & 2 & 7803.49747 & 0.00020 & 0.672 & 0.025 & 41.42 & 0.59\\
  & 216 & 2 & 7805.91882 & 0.00017 & 0.652 & 0.020 & 42.21 & 0.48\\
  & 217 & 2 & 7808.34123 & 0.00023 & 0.735 & 0.035 & 41.98 & 0.71\\
  & 218 & 2 & 7810.76273 & 0.00019 & 0.674 & 0.029 & 42.45 & 0.62\\
  & 219 & 2 & 7813.18456 & 0.00024 & 0.668 & 0.024 & 42.02 & 0.72\\
  & 220 & 2 & 7815.60583 & 0.00017 & 0.725 & 0.024 & 42.88 & 0.55\\
  & 221 & 2 & 7818.02821 & 0.00020 & 0.763 & 0.024 & 42.28 & 0.53\\
  & 222 & 2 & 7820.45019 & 0.00022 & 0.674 & 0.024 & 41.76 & 0.60\\
  & 223 & 2 & 7822.87188 & 0.00021 & 0.756 & 0.028 & 42.67 & 0.66\\
  & 224 & 2 & 7825.29388 & 0.00022 & 0.672 & 0.025 & 42.12 & 0.66\\
  & 225 & 2 & 7827.71513 & 0.00022 & 0.718 & 0.029 & 42.21 & 0.62\\
  & 226 & 2 & 7830.13713 & 0.00026 & 0.705 & 0.033 & 42.90 & 0.79\\
  & 227 & 2 & 7832.55888 & 0.00015 & 0.732 & 0.026 & 42.95 & 0.49\\
  & 228 & 2 & 7834.98120 & 0.00025 & 0.673 & 0.030 & 42.56 & 0.81\\
  & 229 & 2 & 7837.40280 & 0.00017 & 0.697 & 0.024 & 41.98 & 0.52\\
  & 230 & 2 & 7839.82415 & 0.00031 & 0.679 & 0.063 & 42.05 & 0.86\\
\hline
d & -4  & 2 & 7653.94261 & 0.00051 & 0.390 & 0.038 & 49.00 & 2.20\\
  & -3  & 2 & 7657.99220 & 0.00063 & 0.333 & 0.027 & 49.20 & 1.90\\
  & -2  & 2 & 7662.04284 & 0.00051 & 0.395 & 0.030 & 50.10 & 1.50\\
  & -1   & 2 & 7666.09140 & 0.00130 & 0.308 & 0.043 & 48.90 & 2.80\\
  & 0   & 2 & 7670.14198 & 0.00066 & 0.366 & 0.045 & 49.00 & 1.60\\
  & 33  & 2 & 7803.79083 & 0.00049 & 0.366 & 0.021 & 49.40 & 1.30\\
  & 34  & 2 & 7807.84032 & 0.00030 & 0.384 & 0.020 & 49.39 & 0.91\\
  & 35  & 2 & 7811.89116 & 0.00050 & 0.382 & 0.024 & 48.60 & 1.40\\
  & 36  & 2 & 7815.94064 & 0.00030 & 0.348 & 0.019 & 50.21 & 0.85\\
  & 37  & 2 & 7819.99050 & 0.00050 & 0.312 & 0.021 & 49.80 & 1.40\\
  & 38  & 2 & 7824.04185 & 0.00067 & 0.383 & 0.025 & 49.20 & 1.50\\
  & 39  & 2 & 7828.09082 & 0.00043 & 0.387 & 0.031 & 49.40 & 1.10\\
  & 40  & 2 & 7832.14036 & 0.00037 & 0.331 & 0.023 & 48.90 & 1.10\\
  & 41  & 2 & 7836.19171 & 0.00042 & 0.345 & 0.023 & 48.80 & 1.10\\
\hline
e & -1  & 2 & 7654.27862 & 0.00049 & 0.582 & 0.043 & 58.60 & 1.40\\
  & 0   & 2 & 7660.38016 & 0.00078 & 0.495 & 0.047 & 56.40 & 1.90\\
  & 24  & 2 & 7806.75758 & 0.00041 & 0.439 & 0.027 & 55.70 & 1.20\\
  & 25  & 2 & 7812.85701 & 0.00034 & 0.449 & 0.023 & 55.10 & 1.00\\
  & 26  & 2 & 7818.95510 & 0.00030 & 0.482 & 0.022 & 56.07 & 0.92\\
  & 27  & 2 & 7825.05308 & 0.00035 & 0.432 & 0.024 & 55.44 & 0.96\\
  & 28  & 2 & 7831.15206 & 0.00027 & 0.516 & 0.018 & 56.19 & 0.85\\
  & 29  & 2 & 7837.24980 & 0.00025 & 0.499 & 0.021 & 55.58 & 0.71\\
\hline 
f & -2  & 2 & 7652.98579 & 0.00032 & 0.658 & 0.020 & 65.90 & 1.40\\
  & -1  & 2 & 7662.18747 & 0.00040 & 0.620 & 0.037 & 65.30 & 1.80\\
  & 0   & 2 & 7671.39279 & 0.00072 & 0.692 & 0.070 & 64.70 & 4.00\\
  & 15  & 2 & 7809.47554 & 0.00027 & 0.648 & 0.025 & 63.47 & 0.82\\
  & 16  & 2 & 7818.68271 & 0.00032 & 0.634 & 0.022 & 63.35 & 0.93\\
  & 17  & 2 & 7827.88669 & 0.00030 & 0.628 & 0.024 & 63.11 & 0.86\\
  & 18  & 2 & 7837.10322 & 0.00032 & 0.610 & 0.023 & 63.36 & 0.95\\
\hline
g & -1  & 2 & 7652.99481 & 0.00030 & 0.817 & 0.028 & 68.60 & 1.00\\
  & 0   & 2 & 7665.35151 & 0.00028 & 0.691 & 0.026 & 67.49 & 0.82\\
  & 12  & 2 & 7813.60684 & 0.00023 & 0.777 & 0.020 & 68.88 & 0.72\\
  & 13  & 2 & 7825.96112 & 0.00020 & 0.793 & 0.019 & 68.30 & 0.79\\
  & 14  & 2 & 7838.30655 & 0.00028 & 0.695 & 0.026 & 68.08 & 0.98\\
\hline  
h & 0   & 2 & 7662.55467 & 0.00054 & 0.348 & 0.024 & 76.1 & 2.1\\
  & 9   & 2 & 7831.46625 & 0.00047 & 0.346 & 0.016 & 77.2 & 1.3\\
\end{longtable}

\twocolumn

\section{Global analysis: supplementary material}

\subsection{Binned residuals RMS vs. bin size plots}
\label{plotrms}

\begin{figure*}
\centering
\includegraphics[width=0.85\textwidth]{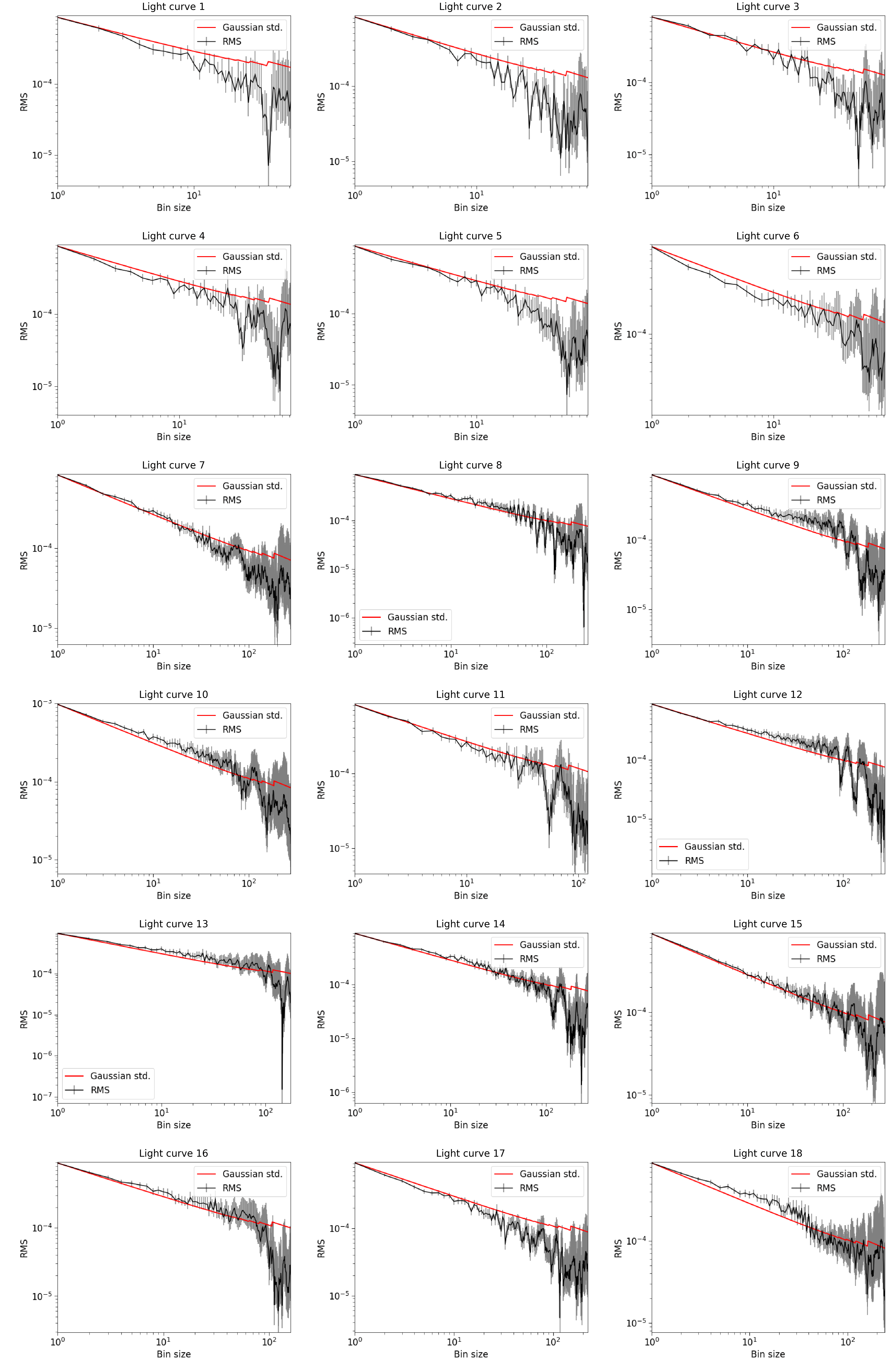}
\caption{Binned residuals RMS (black curves with grey error bars) vs. bin size for the first eighteen \textit{Spitzer}/IRAC light curves. The red curves are the expected RMS for Gaussian noise.}
\label{fig:RMS1} 
\end{figure*}

\begin{figure*}
\centering
\includegraphics[width=0.85\textwidth]{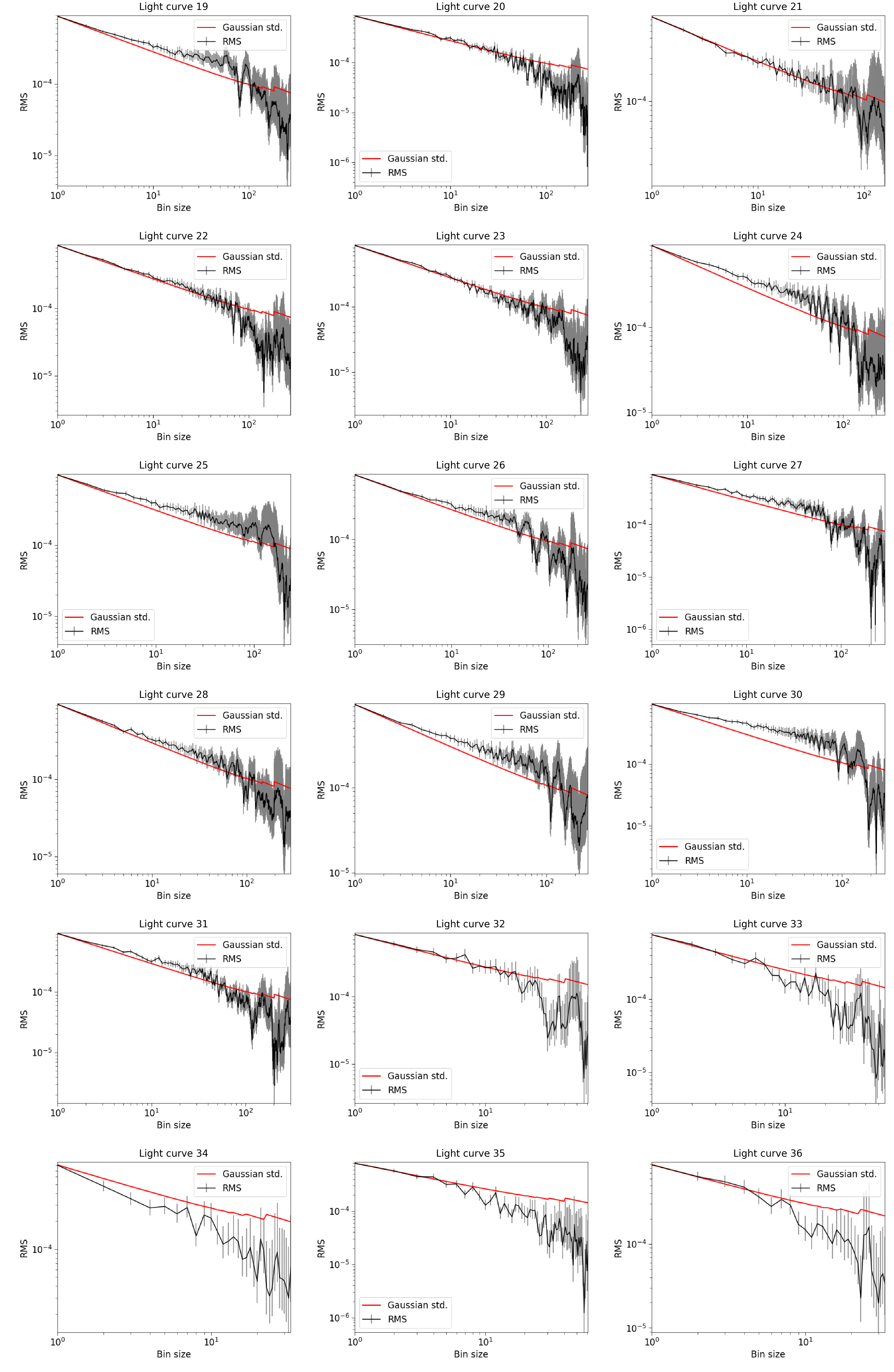}
\caption{Same as Fig. \ref{fig:RMS1}, but for light curves 19 to 36.}
\label{fig:RMS2} 
\end{figure*}

\begin{figure*}
\centering
\includegraphics[width=0.85\textwidth]{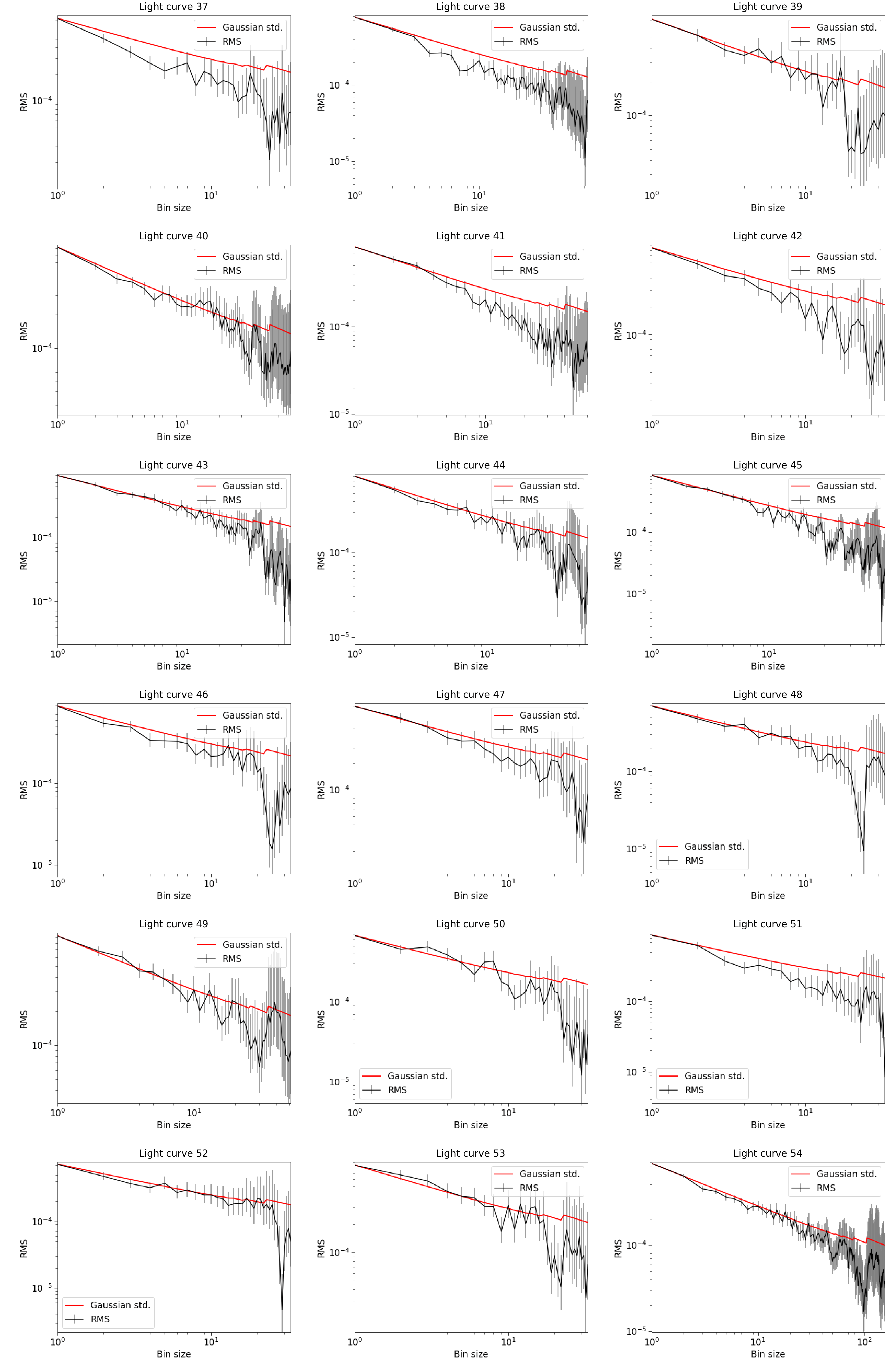}
\caption{Same as Fig. \ref{fig:RMS1}, but for light curves 37 to 54.}
\label{fig:RMS3} 
\end{figure*}

\begin{figure*}
\centering
\includegraphics[width=0.85\textwidth]{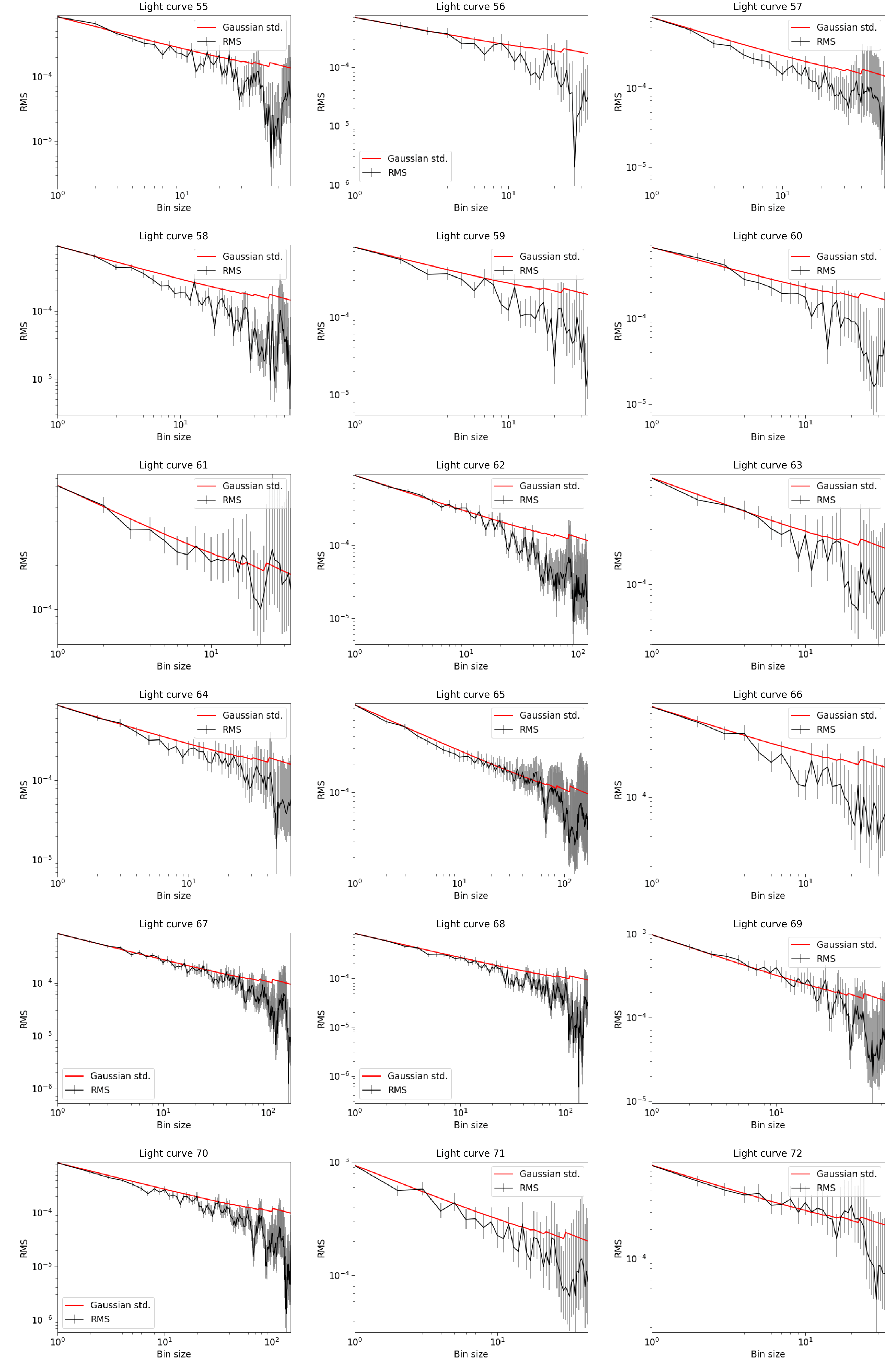}
\caption{Same as Fig. \ref{fig:RMS1}, but for light curves 55 to 72.}
\label{fig:RMS4} 
\end{figure*}

\begin{figure*}
\centering
\includegraphics[width=0.85\textwidth]{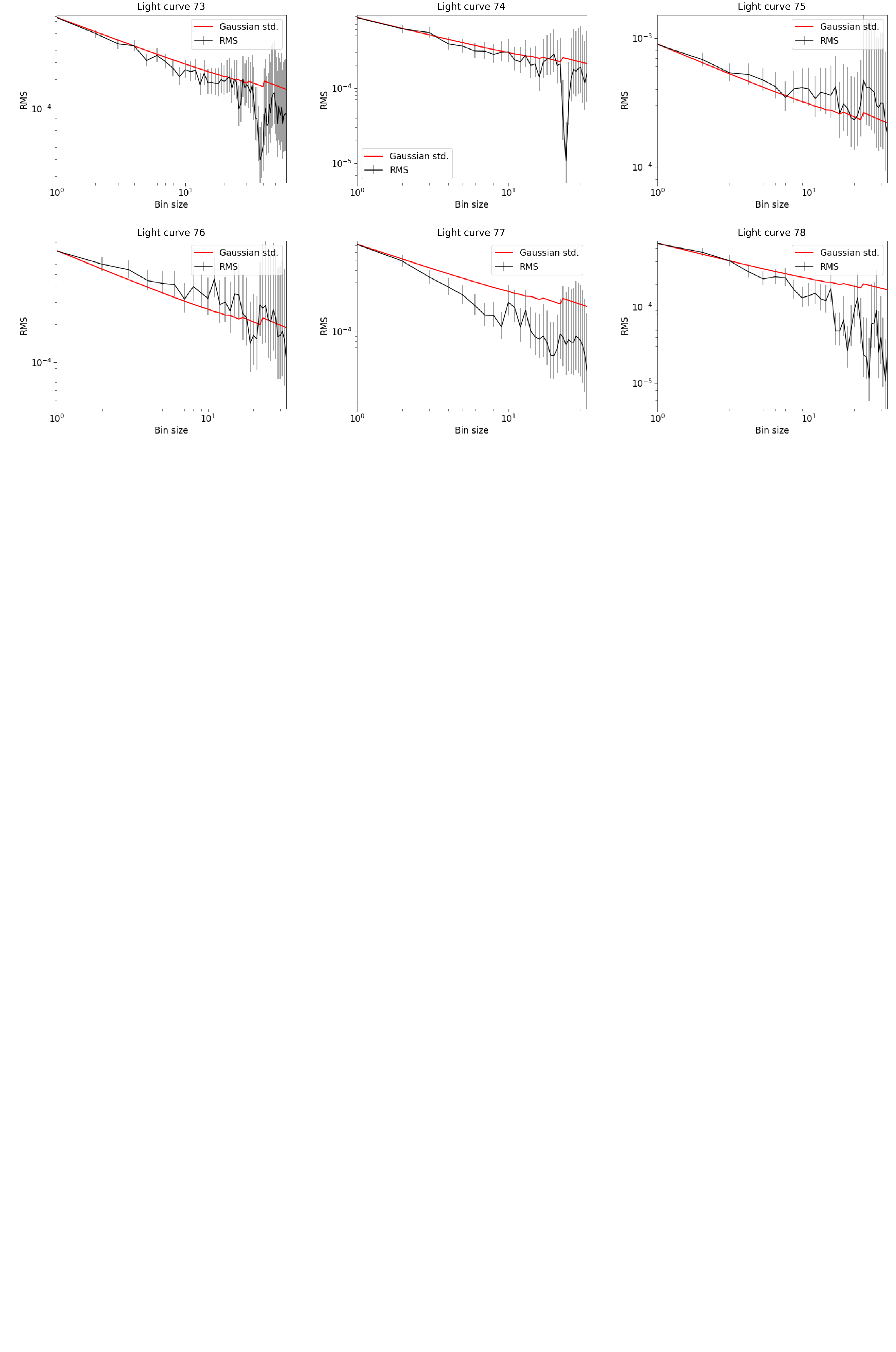}
\vspace{-15cm}
\caption{Same as Fig. \ref{fig:RMS1}, but for light curves 73 to 78.}
\label{fig:RMS5} 
\end{figure*}

Figs.~\ref{fig:RMS1}, \ref{fig:RMS2}, \ref{fig:RMS3}, \ref{fig:RMS4}, and \ref{fig:RMS5} show the RMS vs. bin size plots for the 78 \textit{Spitzer}/IRAC light curves of our dataset, made using the \texttt{binrms} routine of the \texttt{MC$^{3}$} open-source Python package \citep{Cubillos2017}. For each light curve, the RMS of the binned residuals, $\mathrm{RMS}_{N}$, is shown as a black curve for bin sizes (i.e. the mean number $N$ of points in each bin) ranging from one to half the data size. The uncertainty of $\mathrm{RMS}_{N}$ (grey error bars) is computed as $\sigma_{\mathrm{RMS}}=\mathrm{RMS}_{N}/\sqrt{2M}$ (see \citealt{Cubillos2017} for the derivation), where $M$ is the number of bins. The red curve shows the expected RMS $\sigma_{N}$ in the absence of correlated noise, given by \citep{Winn2008}:
\begin{eqnarray}  
\sigma_{N} &=& \frac{\sigma_1}{\sqrt{N}}\sqrt{\frac{M}{M-1}}
\end{eqnarray}
where $\sigma_{1}$ is the RMS of the unbinned residuals. The saw-tooth look of this red curve arises from the
discreet change in $M$, which becomes more significant as $N$ increases.\\
\indent
As mentioned in Section \ref{analyses}, the possible presence of correlated noise in the data is accounted for in our analyses via correction factors $CF$ that we applied to the photometric error bars of each light curve before performing the final analyses. For each light curve, $CF$ is the product of two contributions, $\beta_{w}$ and $\beta_{r}$. On one side, $\beta_{w}$ represents the under- or overestimation of the white noise of each measurement. It is computed as the ratio between the RMS of the unbinned residuals and the mean photometric error. On the other side, $\beta_{r}$ allows to account for possible correlated noise present in the light curve, and is calculated as:
\begin{eqnarray}  
\beta_{r} &=& \frac{\mathrm{RMS}_{N}}{\sigma_{N}}\\ 
&=& \frac{\sigma_N}{\sigma_1}\sqrt{\frac{N(M-1)}{M}}
\end{eqnarray}
The largest value obtained with different bin sizes is kept as $\beta_{r}$.

\newpage
~
\newpage

\subsection{Cross-correlation plots of the posterior PDFs}

\begin{figure}
\centering
\includegraphics[width=0.47\textwidth]{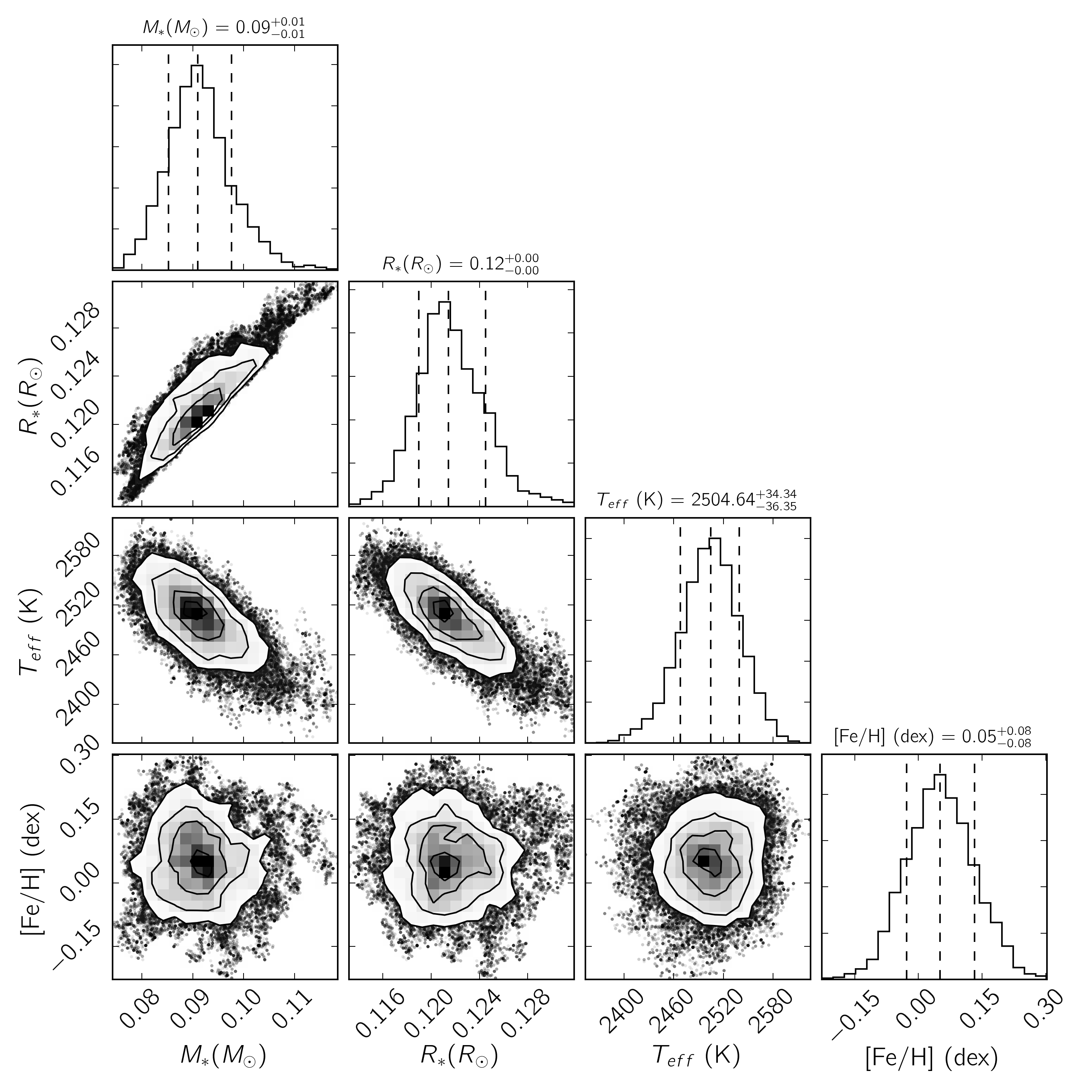}
\caption{Cross-correlation plots and histograms of the posterior PDFs deduced for the stellar parameters from our global analysis. 
}\label{fig:global_correlations1} 
\end{figure}

Fig.~\ref{fig:global_correlations1} shows the cross-correlation of the posterior parameters for the star from the global analysis made using \texttt{corner.py} \citep{corner}. Since the transit duration-period relation constrains the density of the star strongly, the mass and radius are strongly correlated. As the luminosity prior is strongly constrained, the uncertainty in radius is anti-correlated with the effective temperature. The metallicity correlates weakly with the other stellar parameters.

\begin{figure}
\centering
\includegraphics[width=0.98\textwidth]{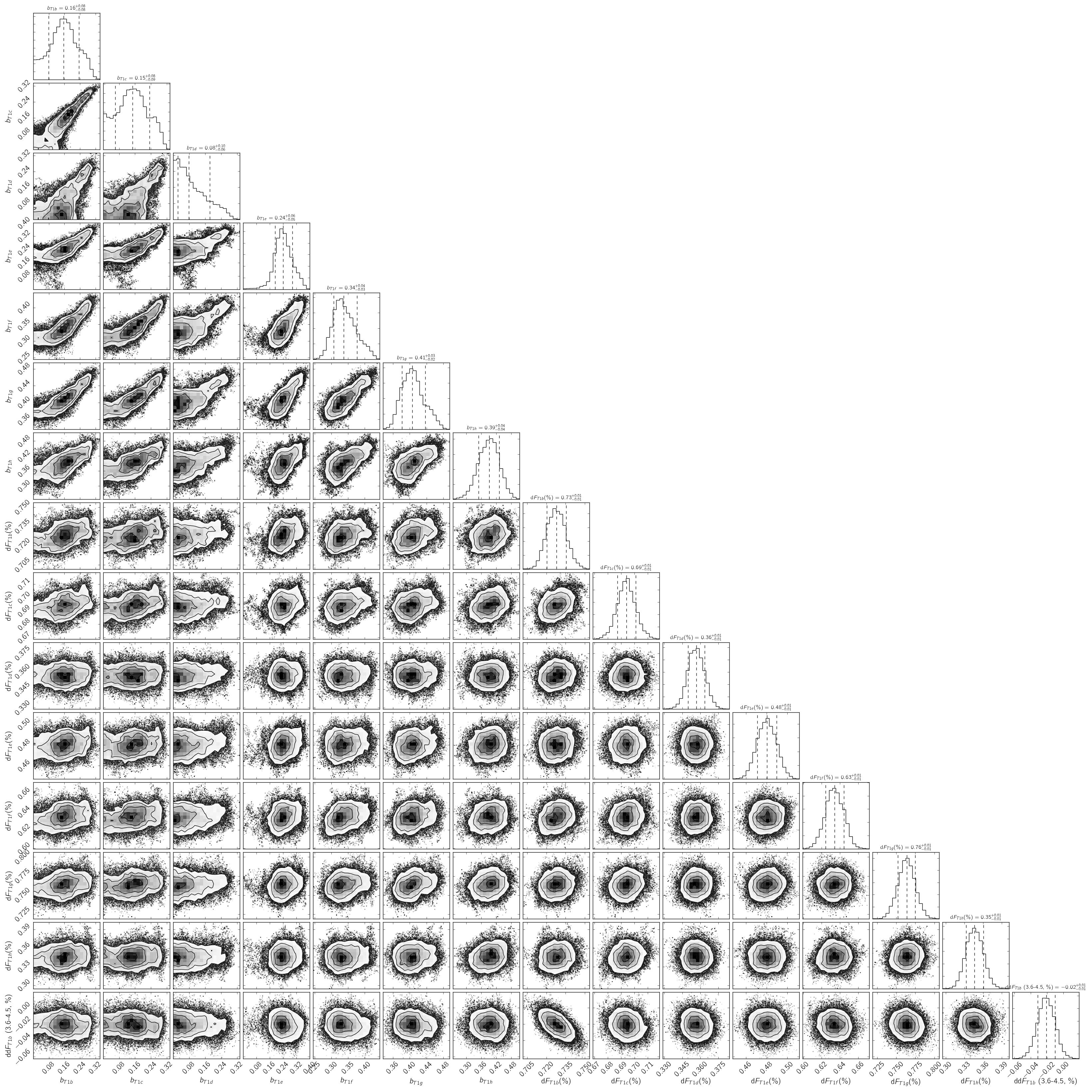}
\caption{Cross-correlation plots and histograms of the posterior PDFs of the transit depths and impact parameters for the seven planets, as well as the difference in transit depth of planet TRAPPIST-1b between IRAC channel 1 and channel 2. 
}\label{fig:global_correlations3} 
\vspace{1cm}
\end{figure}

Fig.~\ref{fig:global_correlations3} shows the correlations between transit impact parameter and depth. As a common density for the star was used for all the transits, the impact parameters are strongly correlated as they anti-correlate with the transit durations, while the durations are well constrained by the data, and as zero eccentricity was assumed for the planet orbits. The depths correlate with the impact parameters as larger impact parameters have transit chords with lower surface brightness (due to limb-darkening), requiring a larger planet to cause the same transit depth.

The limb-darkening parameters and transit times correlate weakly with all of the other parameters, so we have not included these in the cross-correlation plots.

\onecolumn

\onecolumn

\newpage

\subsection{Median values of the residuals in and out of transit}

\begin{longtable}{c | c | c c | c c | c }
\caption{Median values ($\mathrm{median}_{\mathrm{in}}$, $\mathrm{median}_{\mathrm{out}}$) and median absolute deviations ($\sigma_{\mathrm{in}}$ and $\sigma_{\mathrm{out}}$) of the residuals in and out of transit, using the residuals from the global analysis. The last column gives the significance of the difference between $\mathrm{median}_{\mathrm{in}}$ and $\mathrm{median}_{\mathrm{out}}$, computed as $\frac{ \mid\mathrm{median}_{\mathrm{in}}-\mathrm{median}_{\mathrm{out}}\mid}{\sqrt{\sigma_{\mathrm{in}}^2+\sigma_{\mathrm{out}}^2}}$.}\label{tab:discuss}\\
\hline
\hline
Planet & Epoch & $\mathrm{median}_{\mathrm{in}}$ & $\sigma_{\mathrm{in}}$ & $\mathrm{median}_{\mathrm{out}}$ & $\sigma_{\mathrm{out}}$ & Significance of \\
 & & [ppm] & [ppm] & [ppm] & [ppm] & the difference [$\sigma$] \\
\hline 
\endfirsthead
\multicolumn{7}{c}{\tablename\ \thetable{} -- \textit{continued from previous page}} \\
\hline
Planet & Epoch & $\mathrm{median}_{\mathrm{in}}$ & $\sigma_{\mathrm{in}}$ & $\mathrm{median}_{\mathrm{out}}$ & $\sigma_{\mathrm{out}}$ & Significance of \\
 & & [ppm] & [ppm] & [ppm] & [ppm] & the difference [$\sigma$] \\
\hline 
\endhead
\hline \multicolumn{7}{c}{\textit{Continued on next page}} \\ 
\endfoot
\hline \hline
\endlastfoot
b (3.6 $\mu$m) & 322 & -188 & 421 & 213 & 600 & 0.55 \\
  & 326 & -371 & 415 & 33 & 488 & 0.63 \\
  & 327 & 449 & 287 & 26 & 620 & 0.62 \\
  & 330 & -9 & 350 & -107 & 522 & 0.16 \\
  & 333 & -107 & 522 & 151 & 588 & 0.33 \\
  & 334 & -27 & 587 & 143 & 350 & 0.25 \\
  & 335 & 158 & 530 & 90 & 416 & 0.10 \\
  & 338 & 445 & 637 & 162 & 761 & 0.29 \\
  & 339 & 82 & 340 & -153 & 334 & 0.49 \\
  & 341 & 295 & 293 & -45 & 421 & 0.66 \\
  & 342 & -407 & 635 & 219 & 579 & 0.73 \\
\hline
b (4.5 $\mu$m) & 78 & 133 & 420 & -290 & 464 & 0.68 \\
  & 86 & -32 & 344 & 362 & 390 & 0.76 \\
  & 93 & -240 & 303 & -192 & 740 & 0.06 \\
  & 218 & -265 & 407 & 350 & 588 & 0.86 \\
  & 219 & 279 & 527 & -318 & 413 & 0.89 \\
  & 220 & -386 & 353 & 194 & 550 & 0.89 \\
  & 222 & -406 & 583 & 90 & 570 & 0.61 \\ 
  & 224 & 104 & 519 & -240 & 659 & 0.41 \\
  & 225 & -328 & 471 & 179 & 612 & 0.65 \\
  & 226 & -136 & 372 & 196 & 480 & 0.55 \\
  & 227 & -512 & 354 & -361 & 621 & 0.21 \\  
  & 228 & 192 & 279 & -188 & 423 & 0.75 \\
  & 229 & -226 & 831 & -296 & 338 & 0.08 \\
  & 230 & -242 & 732 & -684 & 624 & 0.46 \\
  & 231 & 89 & 979 & 122 & 384 & 0.03 \\
  & 318 & -59 & 382 & -430 & 606 & 0.52 \\
  & 320 & 182 & 379 & -67 & 473 & 0.41 \\
  & 321 & -229 & 466 & -27 & 669 & 0.25 \\
  & 324 & 226 & 417 & 14 & 561 & 0.30 \\
  & 325 & 208 & 553 & 199 & 538 & 0.01 \\
  & 328 & -145 & 700 & 110 & 530 & 0.29 \\
  & 332 & -45 & 530 & -259 & 617 & 0.26 \\
  & 336 & 334 & 669 & -41 & 1003 & 0.31 \\
  & 340 & 359 & 770 & 368 & 362 & 0.01 \\
\hline  
c & 70  & 126 & 475 & 11 & 618 & 0.15 \\
  & 71  & 242 & 302 & -171 & 404 & 0.82 \\
  & 152 & -331 & 251 & -645 & 360 & 0.71 \\
  & 153 & 102 & 405 & -85 & 701 & 0.23 \\
  & 154 & 512 & 538 & 75 & 964 & 0.40 \\
  & 155 & 345 & 623 & -11 & 658 & 0.39 \\
  & 156 & -92 & 684 & 53 & 596 & 0.16 \\
  & 157 & -167 & 742 & -400 & 794 & 0.21 \\
  & 158 & -89 & 371 & -311 & 750 & 0.27 \\
  & 159 & 201 & 653 & -168 & 391 & 0.48 \\
  & 160 & -112 & 380 & -113 & 780 & 0.00 \\
  & 215 & 124 & 602 & -264 & 379 & 0.55 \\
  & 216 & 340 & 664 & -165 & 676 & 0.53 \\
  & 217 & -233 & 536 & 259 & 779 & 0.52 \\
  & 218 & 13 & 547 & -82 & 612 & 0.12 \\
  & 219 & 83 & 721 & -79 & 452 & 0.19 \\
  & 220 & -32 & 472 & 96 & 670 & 0.16 \\
  & 221 & -208 & 711 & 18 & 398 & 0.28 \\
  & 222 & 333 & 501 & -126 & 576 & 0.60 \\
  & 223 & -241 & 509 & 286 & 614 & 0.66 \\
  & 224 & 90 & 772 & -43 & 560 & 0.14 \\
  & 225 & -229 & 561 & -79 & 376 & 0.22 \\
  & 226 & -71 & 602 & -179 & 588 & 0.13 \\
  & 227 & -93 & 556 & -297 & 734 & 0.22 \\
  & 228 & 345 & 777 & -98 & 333 & 0.52 \\
  & 229 & 14 & 507 & -128 & 466 & 0.21 \\
  & 230 & 399 & 746 & -115 & 559 & 0.55 \\
\hline
d & -4  & -159 & 754 & 260 & 733 & 0.40 \\
  & -3  & 110 & 472 & 112 & 551 & 0.00 \\
  & -2  & -59 & 849 & -103 & 674 & 0.04 \\
  & -1  & 61 & 593 & -5 & 409 & 0.09 \\
  & 0   & -83 & 538 & -96 & 669 & 0.02 \\
  & 33  & 45 & 580 & -63 & 721 & 0.12 \\
  & 34  & -73 & 511 & 180 & 579 & 0.33 \\
  & 35  & -87 & 723 & 23 & 451 & 0.13 \\
  & 36  & 157 & 428 & 185 & 478 & 0.04 \\
  & 37  & 313 & 652 & -118 & 478 & 0.04 \\
  & 38  & -18 & 727 & 189 & 656 & 0.21 \\
  & 39  & -131 & 320 & 292 & 425 & 0.80 \\
  & 40  & 312 & 720 & 206 & 573 & 0.12 \\
  & 41  & 54 & 617 & 191 & 646 & 0.15 \\
\hline
e & -1  & -128 & 627 & 159 & 763 & 0.29 \\
  & 0   & -316 & 487 & 87 & 540 & 0.55 \\
  & 24  & 218 & 500 & -257 & 446 & 0.71 \\
  & 25  & -257 & 446 & -49 & 693 & 0.25 \\
  & 26  & -149 & 637 & 234 & 559 & 0.45 \\
  & 27  & 234 & 606 & -392 & 508 & 0.79 \\
  & 28  & -201 & 695 & 334 & 580 & 0.59 \\
  & 29  & -251 & 531 & -5 & 736 & 0.27 \\
\hline 
f & -2  & -235 & 635 & 27 & 512 & 0.32 \\
  & -1  & 195 & 440 & 31 & 560 & 0.23 \\
  & 0   & 40 & 632 & 159 & 726 & 0.12 \\
  & 15  & -205 & 516 & -307 & 793 & 0.11 \\
  & 16  & -274 & 641 & 188 & 493 & 0.57 \\
  & 17  & -117 & 571 & -203 & 612 & 0.10 \\
  & 18  & 79 & 621 & -196 & 541 & 0.33 \\
\hline
g & -1  & -235 & 635 & -230 & 525 & 0.01 \\
  & 0   & 555 & 531 & 377 & 557 & 0.23 \\
  & 12  & -48 & 336 & +199 & 735 & 0.31 \\
  & 13  & -165 & 611 & 209 & 334 & 0.54 \\
  & 14  & 143 & 609 & -365 & 501 & 0.64 \\
\hline  
h & 0   & 14 & 459 & -26 & 646 & 0.05 \\
  & 9   & -27 & 540 & -155 & 557 & 0.16 \\
\end{longtable}

\twocolumn


\bsp	
\label{lastpage}
\end{document}